\pdfoutput=1

\documentclass[useAMS,usenatbib,times,letter,amssymb]{mn2e}
\usepackage{graphicx, float, amsmath,epsfig,times,amssymb,color,verbatim, widetext }
\pdfoutput=1
\def\be{\begin{equation}}
\def\ee{\end{equation}}
\def\bea{\begin{eqnarray}}
\def\eea{\end{eqnarray}}
\def\lg10{\log_{10}}

\newcommand{\bm}[1]{\mbox{\boldmath{$#1$}}}   

\title[Mass profile reconstruction with magnification]{Cluster mass profile reconstruction with size and flux magnification on the HST STAGES survey}

\author[C. Duncan et al.]{Christopher A. J. Duncan$^{1}$\thanks{Email: cajd@roe.ac.uk}, Catherine Heymans$^{1}$, Alan F. Heavens$^{2}$, Benjamin Joachimi$^{3}$\\
$^{1}$Scottish Universities Physics Alliance, Institute for Astronomy, University of Edinburgh, Royal Observatory, Blackford Hill, Edinburgh, EH9 3HJ, UK\\
$^{2}$Imperial Centre for Inference and Cosmology, Department of Physics, Imperial College London, Blackett Laboratory, Prince Consort Road, London, SW7 2AZ, UK.\\
$^{3}$Department of Physics and Astronomy, University College London, Gower Place, London, WC1E 6BT, UK.}

\par

\begin{document}

\date{Accepted 1 January 2016}

\pagerange{\pageref{firstpage}--\pageref{lastpage}} \pubyear{2015}

\maketitle

\label{firstpage}

\begin{abstract}
We present the first measurement of individual cluster mass estimates using weak lensing size and flux magnification.  Using data from the HST-STAGES survey of the A901/902 supercluster we detect the four known groups in the supercluster at high significance using magnification alone.  We discuss the application of a fully Bayesian inference analysis, and investigate a broad range of potential systematics in the application of the method. We compare our results to a previous weak lensing shear analysis of the same field finding the recovered signal-to-noise of our magnification-only analysis to range from 45\% to 110\% of the signal-to-noise in the shear-only analysis.  On a case-by-case basis we find consistent magnification and shear constraints on cluster virial radius,  and finding that for the full sample, magnification constraints to be a factor $0.77\pm 0.18$ lower than the shear measurements. 
\end{abstract}


\section{Introduction}

Galaxy clusters comprise the largest known gravitationally bound objects in the Universe. They can give information on the formation of structure and the cosmological model through knowledge of the underlying density field. In order to interpret galaxy clusters in a cosmological scenario, one must have knowledge of the individual masses of the clusters that enter into the sample. Many different observables are commonly used as a proxy for cluster mass, including cluster member counts (cluster richness) which rely on discrete observable sources as tracers of the underlying matter distribution, or X-ray luminosity and temperature and the Sunyaev-Zeldovich effect which utilise the effects of hot gas in the vicinity of the cluster. In each of these cases, one must make simplifying assumptions about how these tracers follow the underlying dominant dark matter distribution, and take this into account when interpreting the measurement as a proxy for the mass of the cluster. For the use of cluster members in the optical this requires knowledge of the galaxy bias. For X-ray derived masses one must assume hydrostatic equilibrium, although recent studies suggest that X-ray derived masses may be biased low \citep{Simet:2015p2852}. By contrast, gravitational lensing uses measurements of background galaxy size, shape or luminosity to probe the lensing total matter distribution, and is insensitive to the nature of the lensing matter itself.

The use of gravitational lensing measurements as a method of mass reconstruction to date have predominantly dealt exclusively with the shape distortion of distant sources and as a result much time has been invested in developing the tools to accurately use shear measurements. As an example, competitive analyses of the accuracy and precision of weak lensing observable measurement, such as the STEP \citep{2006MNRAS.368.1323H,2007MNRAS.376...13M} and GREAT \citep{Bridle:2009p1030,2010arXiv1009.0779K,2014ApJS..212....5M} programs, have primarily focussed their attention on testing the ability of particular algorithms in measuring source ellipticity with estimates of source size a secondary concern. Ideally, one would like to utilise the maximum number of probes in weak lensing analyses, as a means of reducing the statistical errors on measurements for a given source sample, but also as a means of mitigating systematics in each individual analysis. 

There has been a recent increasing trend to investigate the use of other weak lensing observables, including numerous convincing detections of fluctuations in source counts due to lensing by foreground matter, most frequently dubbed `flux magnification' or `magnification bias'. These analyses measured angular correlation functions between radially separated bins \citep{Myers:2003p2024,Scranton:2005p1124,Hildebrandt:2009p845,Morrison:2012p1286}, and around stacked foreground over-densities as a means of measuring stacked mass profiles \citep{Ford:2014p2751,Ford:2014p2825,Bauer:2011p2066,Umetsu:2014p2726,Hildebrandt:2011p2755} or determining dust profiles \citep{Menard:2010p1495,Hildebrandt:2013p2756}. Of particular note, the analyses of \citet{Hildebrandt:2011p2755,Hildebrandt:2013p2756} measure the mass profiles for high redshift lenses, using a high redshift background source sample where shape determination would be expected to fail, and thus utilises one of the main strengths of a number-counts magnification analysis.   

Contemporaneously, there have been a series of theoretical investigations into the use of the magnification signal to measure cosmological parameters, through the clustering of a photometric sample in \cite{Duncan:2014p2569,Joachimi:2010p855,Waerbeke:2010p7,2010MNRAS.401.2093V} or as part of a joint analysis using a photometric and spectroscopic sample \citep{Gaztanaga:2012p1194,Eriksen:2015p2849}. It is generally found that whilst magnification alone is uncompetitive with shear when an unknown galaxy bias must be simultaneously measured with the data, the combination of clustering and shear can give a significant increase in constraining power over the shear-only signal through degeneracy lifting between the clustering, shear and galaxy-galaxy lensing. Further, \cite{Joachimi:2010p855} found that the addition of existing number density information to a shear analysis on a photometric sample can successfully counteract the loss of information due to the marginalisation over a flexible intrinsic alignment model. Such a combined analysis was adopted as part of the primary science driver in Euclid \citep{Laureijs:2011p2021}, however in \cite{Duncan:2014p2569} it was shown that systematic uncertainties in the magnification signal can lead to catastrophic biases in cosmological model parameters.

Similarly, there has been a recent uptake in investigations into the direct use of size and magnitude measurements to infer lensing properties, either through a comparison in statistics of lensed samples to unlensed samples \citep{Heavens:2013p1550,Alsing:2014p2846,Casaponsa:2013p1480}, or through the use of the fundamental plane relation \citep{Huff:2011p1392,Sonnenfeld:2011p1035,Bertin:2006p2063} . In each case, major astrophysical systematics, similar to intrinsic alignments for a shear analysis, may be present through intrinsic size-density correlations \citep{Ciarlariello:2014p2844}, or the correlation between fundamental plane residuals and density \citep{2015arXiv150402662J}. 

In \cite{Heavens:2013p1550} it was demonstrated that substantial gains could be made in the combination of size magnification with shear, particularly when noise dominated, and noted that the noise-free size measurement can be made to be uncorrelated to the shear measurement provided that the size is measured as the square-root of a measured source area. 

\cite{Rozo:2010p1496} forecast an improvement of $\sim 50\%$ in cluster mass estimates from a joint size- magnification, clustering and shear analysis over shear-only. \cite{Eifler:2013p2722} found that constraints on a set of cosmological parameters from a non-tomographic COSEBI shear analysis were significantly improved with the addition of projected clustering information, but that the further inclusion of direct magnification did not give significant further improvement. 

In \cite{Alsing:2014p2846} the authors forecast using a theoretically motivated linear alignment and intrinsic-size-density correlation model that the combination of size and magnitude magnification with shear can give improvements in dark energy parameters of $\sim 25\to65\%$, whilst quantifying the typical dispersion on the inferred convergence field using an intrinsic size-magnitude distribution measured with CFHTLenS. 

In \cite{Casaponsa:2013p1480} it was shown through the use of image simulations that size measurements using \emph{lens}Fit \citep{Miller:2007p2375} could estimate the convergence field in an unbiased way provided the source sample was selected to be above a flux signal--to--noise ratio of $10$, and the galaxies are larger than the point spread function. They concluded that high resolution space-based imaging is ideal for a size-magnification analysis.

A recent observational application of the use of the size and magnitude magnification effect is the application in \cite{Schmidt:2012p1106} to stacked group lensing in the COSMOS field. In this paper, authors claim a detection of the magnification effect with a signal--to--noise ratio of $\sim 40\%$ of the shear using a maximum-likelihood estimator based around the assumption of log-normality in the size distribution and Gaussianity in the magnitude distribution. In this paper, we instead apply our method of mass estimation using galaxy sizes and magnitudes to individual large clusters of $M=  O(10^{14}) M_{\odot}/h$ in the STAGES super cluster.

In section \ref{sec:TheoryAndMethod} we detail relevant weak lensing theory, and detail a Bayesian method for determining cluster model parameters for a given lens from source size, magnitude and ellipticity measurements whilst avoiding some of the simplifying assumptions of previous analyses. We discuss how a joint analysis using all three observables could be combined in a self-consistent way. In Section \ref{sec:STAGESDataset} we describe the STAGES dataset and selection of the source sample. In Section \ref{sec:ApplicationToMocks}, the method is applied to mock catalogues designed to reflect the main features of the data-set, and conclusions are drawn on the ability to utilise the method to measure cluster model parameters on different mass lenses, and quantify the effect of limitations in the data-set and simplifying assumptions. Finally, in Section \ref{sec:ApplicationToSTAGES} the method is applied to the STAGES dataset, and results are presented for the STAGES clusters and compared to pre-existing shear measurements. We conclude in Section \ref{Sec:Conclusions}. 

Throughout this paper we assume a flat fiducial cosmology with $w = -1$, $\Omega_M = 0.3$, $\Omega_\lambda = 0.7$ and $h = 0.7$. Magnitudes are given in the AB system.

\section{Theory and Method}\label{sec:TheoryAndMethod}
\subsection{Weak Lensing Theory}

As photons propagate past a foreground matter density contrast, its path is deflected according to the Jacobian mapping between the source plane and the observed sky as
\be
\mathcal{A} = (1-\kappa) \begin{pmatrix} 1 -g_1 & -g_2 \\ -g_2 & 1+g_1  \end{pmatrix},
\ee
in the linear limit. The convergence ($\kappa$) and  complex reduced shear ($g = g_1+ig_2 = \gamma/[1-\kappa]$) vary with angular position on the sky and are functions of gravitational potential of the lens and geometry of the lens-source system. Both the convergence and the shear ($\gamma$) can be related to the projected surface mass density of the lensing matter as
\bea
\kappa(\xi) &=& \Sigma_{\rm Crit}^{-1}\Sigma[\xi], \\
\gamma(\xi) &=& \Sigma_{\rm Crit}^{-1}[\langle \Sigma \rangle (<\xi) - \Sigma(\xi)],
\eea
where $\xi$ is the distance between the source and lens centre on the source plane, and the mean surface mass density within $\xi$ is given by $\langle \Sigma \rangle (<\xi)$. The critical surface mass density is given by
\be
\Sigma_{\rm Crit} = \frac{c^2}{4\pi G}\frac{D_s}{D_d D_{ds}},\label{eqn:Sigma_Crit}
\ee
where $D_s$ and $D_d$ is the angular diameter distance to the source and lens, and  $D_{ds}$ is the angular diameter distance between the source and lens.

 The convergence denotes an isotropic stretching of the source image, with a corresponding change in the observed size of the source. As a result of the applicability of Liouville's Theorem, this change in source size corresponds directly to a change in the observed flux of the source. Consequently, the lensed size and flux of a source can be related to its unlensed quantities according to 
\bea
R &=& \mu^{\frac{1}{2}}R_0, \label{eqn:Lensing_Relations__Size}\\
S &=& \mu S_0, \label{eqn:Lensing_Relations__Flux}\\
m &=& m_0 +2.5\lg10{\mu}, \label{eqn:Lensing_Relations__Magnitude}
\eea
where $R$, $S$ and $m$ represent the source size\footnote{The source size is typically defined as the square-root of the area of the source}, flux and magnitude respectively, subscript ``$0$'' denotes intrinsic (or unlensed) quantities, and the local magnification factor $\mu$ is given by
\be
\mu = [\det(\mathcal{A})]^{-1} = [(1-\kappa)^2 - \gamma^2]^{-1}.
\ee
The action of a magnification field is therefore to alter the size and brightness of a lensed source, or locally shift the size-magnitude distribution for the source sample. Equivalently, one may consider the action of the magnification field as a local shift in the imposed source size and flux limits of the analysis or data: together with changes in the observed position of the sources, this forms the basis of flux-magnification analyses through clustering statistics. 

\begin{figure*}
\centering
\includegraphics[width = 0.8\textwidth]{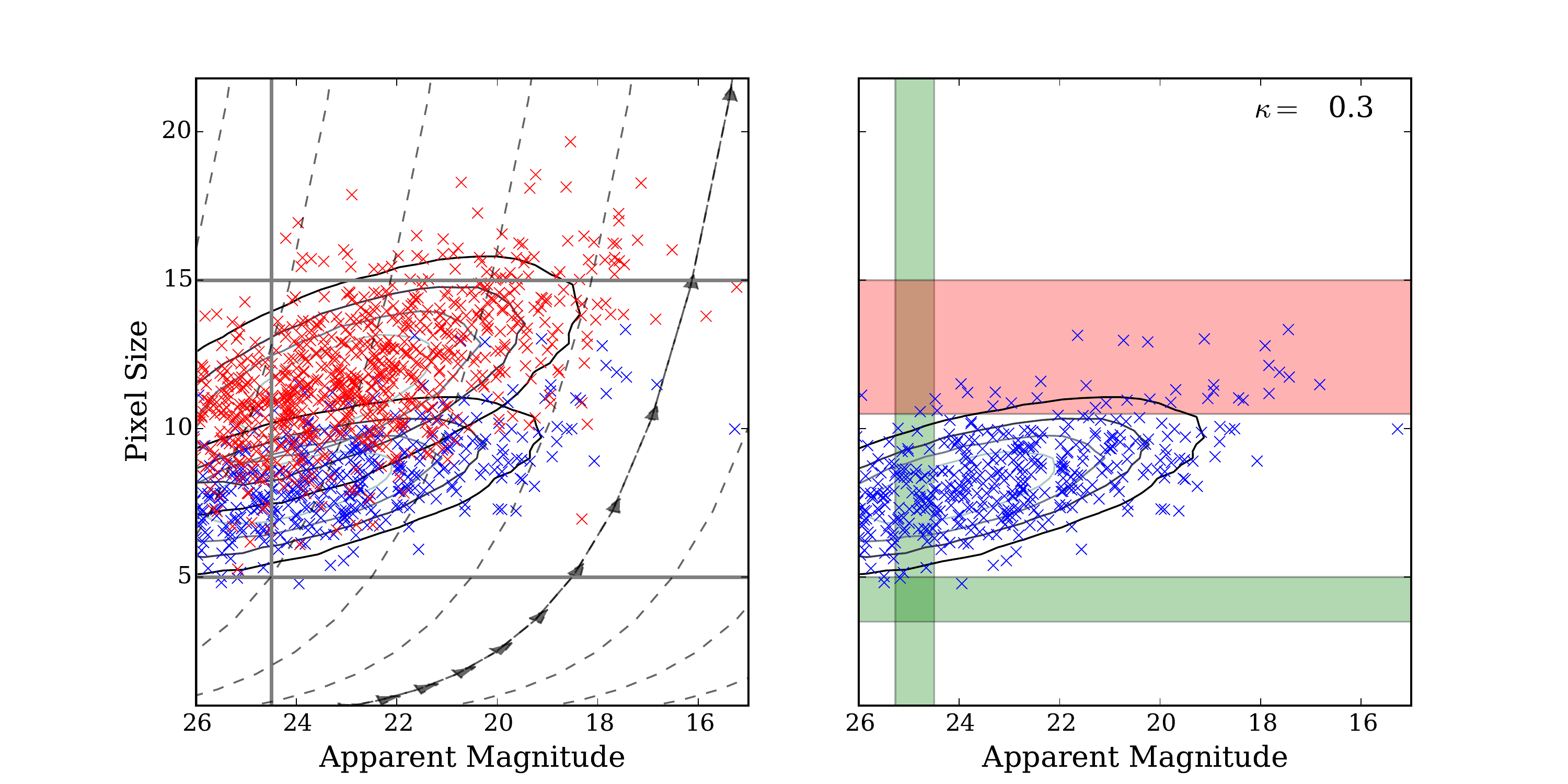}
\caption{Illustrative figure showing the effect of lensing on a body of sources whose sizes and magnitudes are sampled from a multivariate Gaussian. Blue crosses correspond to unlensed sources, whilst red crosses ({\it left} panel only) show lensed counterparts after a constant convergence field of $\kappa = 0.3$ is applied. Dashed lines (and arrows) show the direction of shift in the size-magnitude parameter plance after the application of the convergence field. Horizontal and vertical lines show limits on the observed source size and magnitude respectively. The {\it right} panel shows the equivalent local change in enforced source size and magnitude limits: red areas show regions of parameter space now unobservable on the lensed patch of sky, whilst green areas show regions only observable due to the action of the local convergence field. Sources in the red(green) patch are therefore removed(added) to the observed source sample.}\label{fig:Size_Mag_Lensing_Example}
\end{figure*}

Figure \ref{fig:Size_Mag_Lensing_Example} shows as an example the action of a constant positive convergence field (associated with a lensing foreground over-density) on a model size-magnitude distribution in the presence of a bright magnitude limit, and large and small size limits. The action of such a field is to make the observed sources larger and brighter than their intrinsic values (blue crosses to red on the left panel), consequently locally removing or adding sources to the sample (red and blue regions in the right panel).

\subsection{Bayesian Mass Profile Reconstruction}\label{sec:Reconstruction_Method}

\subsubsection{Motivation}
In \citet[][]{Heavens:2013p1550,Alsing:2014p2846,Casaponsa:2013p1480,Schmidt:2012p1106} the authors presented the framework for the use of a frequentist estimator based method of probing the magnification field in differing contexts. Generally, in such an analysis, one constructs an estimator based on the magnification relations given in equations \ref{eqn:Lensing_Relations__Size} to \ref{eqn:Lensing_Relations__Magnitude}. For example, for the size information one can construct an estimator as
 \be
\hat{\mu} =  \left(\frac{R}{\langle R \rangle}_{\rm field}\right)^2
\ee
where the numerator corresponds to the size of the source or the mean of a locally selected source sample, and the denominator corresponds to the mean size over the whole field, assumed to be an unbiased estimator of the mean of the distribution of intrinsic sizes for the sample considered. 

The use of such an estimator requires special care. Firstly, one must take into account the presence of size of flux/magnitude cuts requires an alteration of the relations in equations \ref{eqn:Lensing_Relations__Size} to \ref{eqn:Lensing_Relations__Magnitude} using magnification `responsivity' factors to account for sources being boosted outside these limits, and these factors must be themselves estimated from the data \citep[see ][ for further discussion]{Alsing:2014p2846,Schmidt:2012p1106}. Secondly, the estimator relies on the assumption that the field mean ($\langle R \rangle_{\rm field}$ in this example) is representative of the unlensed mean of the source sample. This can occur when the `field' sample is chosen over an area where the average magnification is not unity, and can be avoided by calculating the field mean over a large area or on a blank field. Thirdly, where a single source is considered, or the source sample is chosen within flux of size ranges, any intrinsic size-luminosity relation must be considered to account for the flux-lensing of the sample, and to ensure that the estimator compares mean sizes of equivalent samples. Finally, such an estimator gives an estimate for the average magnification factor for the source sample. It's physical interpretation is therefore only straight-forward where the sources are selected locally, or on a region where they are expected to experience the same magnification, such as in an annulus around a spherically-symmetric lens mass distribution.

This paper motivates a departure from such a formalism, and in the next section we detail a Bayesian interpretation of the magnification field similar to that detailed in \cite{Alsing:2014p2846}, but with an emphasis on inferring the mass model parameters for an individual lens assuming knowledge of the intrinsic size-magnitude and redshift distributions of the source sample. We discuss in detail the advantage of such a method, and extend it to include ellipticities, as well as discussion the application of a full joint shear and magnification analysis within this framework.

\subsubsection{A joint size and flux magnification analysis}

Consider a single observation of the size and magnitude ($R, m$) of a lensed source, from which we want to place constraints on the mass profile of the lensing medium. In Bayesian nomenclature, we wish to construct a posterior distribution for a set of parameters which define the lensing cluster mass profile (hereafter denoted using $\alpha$) from an observation of lensed quantities. Applying Bayes' theorem, this can be formulated as
\be\label{eqn:BayesTheorem}
p(\alpha|R, m) = \frac{p(R, m|\alpha)p(\alpha)}{p(R,m)} \propto p(R,m|\alpha)p(\alpha).
\ee
The likelihood $[p(R, m|\alpha)]$ describes the probability of making such an observation given a model for the lensing mass profile, and prior knowledge on the cluster mass profile may be set using $p(\alpha)$. For the remainder of this discussion we assume a flat prior, and the evidence $[p(R,m)]$ is taken as a normalising constant, however this can be easily relaxed. 

The likelihood can be related to intrinsic quantities by marginalising over these quantities as nuisance parameters
\bea
p(R, m|\alpha) &=& \int {\rm d}m_0\;{\rm d}R_0\;{\rm d}z\; p(R, m|\alpha, R_0, m_0, z)\nonumber\\
&\times&p(R_0, m_0, z|\alpha),\\
& = & \int {\rm d}m_0\;{\rm d}R_0\;{\rm d}z\; p(R, m|\alpha, R_0, m_0, z)\nonumber\\
&\times&p(R_0, m_0|\alpha)p(z|R_0,m_0,\alpha). \label{eqn:Likelihood_FirstExpansion}
\eea
By integrating over an assumed redshift distribution where the source redshift is not known, the method automatically takes into account the possibility the source lies radially close-to or in front of the lens. The intrinsic size, magnitude and redshift ($R_0, m_0, z$) of the source are taken to be independent of the lensing foreground so $p(R_0, m_0|\alpha) \to p(R_0, m_0)$ and $p(z|R_0,m_0,\alpha) \to p(z|R_0,m_0)$.  By enforcing this simplification, one assumes that there are no intrinsic size-, magnitude- nor redshift-density correlations which could cause a general change in size or magnitude of a population of sources physically close to the lens. This assumption should give accurate results if the source sample is selected to be radially distant from the lens so that the lensing effect dominates, however such separation is not always possible. The implications of such correlations is taken to be outwith the scope of this work, however one may note that given a suitable model for this relation, one can naturally incorporate this model into the intrinsic size-magnitude relation by keeping the $\alpha$ dependence of this term explicit. 

Where the intrinsic size, magnitude and redshift of the source are known, the final line of equation \ref{eqn:Likelihood_FirstExpansion} is described by a product of Dirac Delta functions centred on these values. In this case the magnification factor associated with that lens-source system is well known. In practice, such quantities are not observable, and one may instead marginalise over the distribution of true properties conditioned on observed values. This distribution must be representative of the source sample considered, and therefore accurately reflect the selection criteria in producing the source sample being considered to ensure parameter values are unbiased: for example, where the sample is considered in a tomographic redshift bin, the redshift distribution should reflect this choice. The extension to tomographic samples is trivial, however this comes with the caveat that the formalism presented here assumes that the redshift distribution is that of the {\it true} redshift for the sample: where an uncertainty is associated with the measured redshift, this can be incorporated by integrating over a latent variable (discussed further in section \ref{Sec:Includ_Measurement_Noise}).

In the absence of measurement noise, the former term in equation \ref{eqn:Likelihood_FirstExpansion} contains information on the lensing of the source and can be determined using the relations given in Equations \ref{eqn:Lensing_Relations__Size} to \ref{eqn:Lensing_Relations__Flux} as
\bea\label{eqn:Observed_Intrinsic_SizeMagnitude_Relation}
p(R, m|\alpha, R_0, m_0, z) &=& \delta_D(R - R_0\mu^{\frac{1}{2}}[\alpha,\xi,z])\\
&\times&\delta_D(m - m_0 +2.5\log_{10}\{\mu[\alpha,\xi,z]\}),\nonumber
\eea
where $\xi$ denotes the physical transverse separation of the lens and source and is suppressed for the remainder of this text for clarity. Using a change in variables, the marginalisation over the intrinsic size and magnitude can be carried out so that the likelihood takes the form
\be\label{eqn:Size-Magnitude_SMD_Posterior}
\begin{split}
p(R, m|\alpha) =  &\int \;{\rm d}z \mu^{-\frac{1}{2}}p_{[R_0,m_0| z]}\left(\mu^{-\frac{1}{2}}R,m+2.5\log_{10}\mu\right)\\
& \times \; p_{[z|m_0, R_0]}(z| m+2.5\log_{10}\mu, \mu^{-\frac{1}{2}}R),
\end{split}
\ee
where the notation $p_{[x]}(y)$ denotes the probability density function of $x$ evaluated at $x=y$. The likelihood for each galaxy is then constructed by sampling the intrinsic size-magnitude distribution along a `de-lensing' line, i.e. taking the probability that the source has an intrinsic size and magnitude given by its measured quantities corrected for the modelled local magnification field given by cluster parameters $\alpha$. A similar result is given in equation 9 of \cite{Alsing:2014p2846} where the likelihood is constructed for the convergence assuming the linearisation of the lensing relations.

The posterior on lens mass profile parameters can then be constructed for a single source by reapplication of Bayes' Theorem (as in equation \ref{eqn:BayesTheorem}), and joint constraints using the whole source sample can be obtained by multiplying single-source likelihoods (or summing log-likelihoods) in the usual way.

\subsubsection{Normalisation of the likelihood}

If the source sample is chosen using some selection based on parameters altered by the magnification field (e.g. size, magnitude or flux signal--to--noise) this must be taken into account in the evaluation of the likelihood to avoid inaccurate parameter measurements. In such a case, the application of a non-zero magnification factor will shift the true underlying intrinsic size-magnitude distribution in the size and magnitude planes, altering the normalisation of the likelihood (see Figure \ref{fig:Size_Mag_Lensing_Example}). Where hard size and magnitude cuts are used the likelihood must be normalised such that
\be
\int^{m_u}_{m_l} {\rm d} m \int^{R_u}_{R_l} {\rm d}R \; p(R,m|\alpha) =  1,
\ee
where the integrals are understood to extend over {\it lensed} quantities, between lower and upper limits denoted by subscript $l$ and $u$ respectively. By substituting the form of the likelihood in equation \ref{eqn:Size-Magnitude_SMD_Posterior} and assuming an deterministic relationship between the measured size and magnitude and their unlensed counterparts, the magnification-dependent nature of the normalisation can be made more explicit:
\begin{widetext}
\bea
\int {\rm d}z\; \mu^{-\frac{1}{2}} \int_{m_l}^{m_u} {\rm d}m\int_{R_l}^{R_u} {\rm d}R\; p_{[R_0, m_0|z]}\left(\mu^{\frac{1}{2}}R,m+2.5\log_{10}\{\mu\}\right) p_{[z|m_0,R_0]}\left(z|m+2.5\log_{10}\{\mu\}, \mu^{\frac{1}{2}}R\right), \nonumber\\
= \int {\rm d}z\; \int_{m_l+2.5\log_{10}\{\mu\}}^{m_u+2.5\log_{10}\{\mu\}} {\rm d}m_0\int_{\mu^{-\frac{1}{2}}R_l}^{\mu^{-\frac{1}{2}}R_u} {\rm d}R_0\;p\left(R_0,m_0\right)p(z|m_0, R_0) = 1. \nonumber
\eea
\end{widetext}
The normalisation varies with magnification factor, and consequently with the set of cluster mass profile parameters ($\alpha$) for a given source. In contrast to the case where no cuts are applied, such a normalisation will change the shape of the recovered likelihood, and thus neglecting this effect will bias recovered cluster profile parameters.

Here, we have considered only hard cuts on the data, however in reality it may often be the case that a smooth selection function is applied to the data. Such a case is considered in more detail in \cite{Alsing:2014p2846} and can be easily extended to the analysis presented here where the form of the selection function is known.

\subsubsection{Analysis using sizes or magnitudes only}\label{sec: Size_Magnitude_Only_method}
 
Where only reliable magnitude information is available, posteriors may be produced by marginalising the likelihood given in equation \ref{eqn:Size-Magnitude_SMD_Posterior} over the full range of source sizes considered in the sample
\begin{flalign}
p(m|\alpha) &= \int {\rm d}z \int_{\mu^{-\frac{1}{2}}R_l}^{\mu^{-\frac{1}{2}}R_u} \frac{{\rm d}R_0}{\mu^{\frac{1}{2}} }p_{[R_0, m_0]}\left(R_0,m+2.5\log_{10}\{\mu\}\right)\nonumber\\
&\times p_{[z|m_0,R_0]}(z|m+2.5\log_{10}\{\mu\},R_0),
\end{flalign}
and
\be 
\int^{m_u+2.5\log_{10}\{\mu\}}_{m_l+2.5\log_{10}\{\mu\}} {\rm d}m_0 \; p(m_0|\alpha) = 1.
\ee
In the final relation, we have again assumed a deterministic, lensing-only relation between observed magnitude and intrinsic magnitude, to make the magnification-factor-dependent nature of the normalisation explicit.

Similarly, a size-only likelihood may be formed by marginalising over the lensed magnitude, giving 
\bea\label{eqn:Size-Only_SMD_Posterior}
p(R|\alpha) &=& \int {\rm d}z\; \mu^{-\frac{1}{2}} \int_{m_l+2.5\log_{10}\{\mu\}}^{m_u+2.5\log_{10}\{\mu\}} {\rm d}m_0\;\\
&\times& p_{[R_0, m_0]}\left(\mu^{-\frac{1}{2}}R,m_0\right)p_{[z|m_0,R_0]}(z|m_0, \mu^{-\frac{1}{2}}R),\;\;\;\;\nonumber
\eea
with 
\be
\int^{\mu^{-1/2}R_u}_{\mu^{-1/2}R_l} {\rm d}R_0 \; p(R_0|\alpha) = 1.
\ee

\subsubsection{Extension to ellipticities}

Where ellipticity information is also available, the above formalism can be extended to construct a joint shear and magnification analysis of the lens mass profile. The likelihood can be constructed by integrating over intrinsic quantities as nuisance parameters
\bea
p(R, m, e|\alpha) &=& \int {\rm d}z \; {\rm d}R_0 \; {\rm d}m_0 \; {\rm d}^2e_0 \; p(R, m, e|\alpha, R_0, m_0, e_0, z)\nonumber\\
&\times& p(R_0,m_0,e_0)p(z|R_0,m_0,e_0),
\eea
where $e$ denotes the set of both ellipticity components in a given co-ordinate frame. As before, the second term gives the redshift distribution of the population from which the source was a sampled, and any redshift dependence of the intrinsic ellipticity, size or magnitude can be incorporated into this term. The first term in this equation gives the relation between the observed quantities and the intrinsic quantities, which is assumed to be deterministic and solely due to lensing in the limit of negligible measurement errors

\begin{flalign}
p(R, m, e|&\alpha, R_0, m_0, e_0, z) = \delta_D(R - R_0\mu^{\frac{1}{2}}[\alpha,\xi,z])\nonumber\\
&\times\delta_D(m - m_0 +2.5\log_{10}\{\mu[\alpha,\xi,z]\})\nonumber\\
&\times\delta_D(e - \mathcal{E}[e_0,g]), \label{eqn:Data_to_Instrinsic_Relation_Dirac}
\end{flalign}
where $\mathcal{E}$ denotes the action of the lensing reduced shear on the nuisance intrinsic ellipticity parameter considered, such that $\mathcal{E}^{-1}(e,g) = e_0$ and $\mathcal{E}(e_0,g) = e$.  In the weak lensing limit, the observed ellipticity may be related to the intrinsic ellipticity of the source and the applied shear field by way of a Taylor Expansion
\be
e_{\alpha}(e_0,g) = e^{\alpha}_0 + \frac{\partial e_{\alpha}}{\partial e_{\beta}}\gamma_{\beta} + O(|\gamma|^2) = e^{\alpha}_0 + P^{\gamma}_{\alpha\beta}\gamma_{\beta},
\ee
where the coefficient of the linear term is frequently referred to as the `shear responsivity', and details how the measured ellipticity responds to the applied shear field, and Einstein summation is assumed. In the parlance used here, this can be expressed as
\bea
\mathcal{E}_\alpha = e^{\alpha}_0 + P^{\gamma}_{\alpha\beta}\gamma_{\beta}\nonumber\\
\mathcal{E}^{-1}_\alpha = e^{\alpha} - P^{\gamma}_{\alpha\beta}\gamma_{\beta}\nonumber.
\eea
Similar expressions can be determined where the weak lensing limit has not been applied, as in \cite{Seitz:1995p2763,Seitz:1997p2784}. Using these expressions, the likelihood is then given by
\bea
p(R, m, e|\alpha) &=& \int {\rm d}z \left(\prod_{i=1}^2 \frac{\partial \mathcal{E}^{-1}}{\partial e_{i}}\right) \; p(\mu^{-\frac{1}{2}}R,m+2.5\lg10{\mu},\mathcal{E}^{-1})\nonumber\\
&\times& p(z|\mu^{-\frac{1}{2}}R,m+2.5\lg10{\mu},\mathcal{E}^{-1}).
\eea
When ellipticity measurements only are considered, this can be reduced to 
\be
p(e|\alpha) = \int {\rm d}z  \left(\prod_{i=1}^2 \frac{\partial \mathcal{E}^{-1}}{\partial e_{i}}\right) \; p(\mathcal{E}^{-1})p(z|\mathcal{E}^{-1})
\ee
and the posterior for the source sample constructed as before. 

\subsubsection{Including measurement noise}\label{Sec:Includ_Measurement_Noise} 

So far, we have considered the case where the data is considered exact, however in reality the data will consist of noisy estimators of the true underlying quantity of interest. In this case, the relation between the observed size, magnitude, ellipticity or redshift and the intrinsic values associated with the source galaxies is not longer a deterministic relationship dependent only on the lensing mass, and the relations given in equation \ref{eqn:Data_to_Instrinsic_Relation_Dirac} and \ref{eqn:Observed_Intrinsic_SizeMagnitude_Relation} no longer hold.

Noise in the data can be integrated within this formalism by marginalising over a latent variable which denotes the true lensed quantity. For the source size, magnitude and ellipticity, this requires that
\begin{flalign}
p(R, m, e|\alpha, R_0, m_0, e_0, z) &= \int \; {\rm d}\hat{m}\;{\rm d}\hat{R}\;{\rm d^2}\hat{e} \; p(R, m, e|\hat{R},\hat{m},\hat{e})\nonumber\\
&\times p( \hat{R},\hat{m},\hat{e} | \alpha, R_0, m_0, e_0, z),
\end{flalign}
where variables with a hat denote latent variables which are marginalised over. In this relation, the $p(R, m, e|\hat{R},\hat{m},\hat{e})$ therefore reflects the uncertainty in the measured data, and the latter relation gives the usual lensing relations (given by equation \ref{eqn:Data_to_Instrinsic_Relation_Dirac})).  

Similarly, uncertainty in the redshift estimate can be absorbed into the analysis taking
\be
p(z|R_0, m_0, e_0) \to \int \; {\rm d}\hat{z} \; p(z|\hat{z})p(\hat{z}|R_0, m_0, e_0).
\ee
Where the measurement noise is additive on the quantity of interest, each of these cases considers the convolution of the noise-free likelihood with a distribution describing the uncertainty on the parameter of interest, where the width of the distribution varies with each source. As such, the application of such a marginalisation in brute force will extend the run-time of the likelihood evaluation by a factor of $N$ for each noisy redshift, size or magnitude estimator per source, where $N$ describes the number of times that the noise-free likelihood must be sampled to ensure convergence of the convolution. Where the noise-free likelihood is expensive to calculate (for example due to a large source sample, or the requirement to marginalise over many latent variables), this may result in a prohibitively long run time which requires more advanced techniques to overcome. 

A limitation in the extension to such a marginalisation lies in the fact that the noise-free likelihood has a high dimensionality, as it depends on the source position and cluster model parameters as well as latent size, magnitude, and redshift, so that the evaluation of the noise-free likelihood on a grid which can be applied to all sources (removing this as a bottle-neck) is intractable.  Alternatively, the dependancy on cluster model parameters, source position and redshift can be absorbed into the local magnification factor, thus significantly reducing the dimensionality of the problem and allowing the measurement-noise-free likelihood to be evaluated on a grid of latent lensed size, magnitude and local magnification factor (and redshift if unknown) which can be referenced for each cluster model parameter choice and source considered. Whilst the evaluation of such a grid is expensive where the evaluation of the likelihood per source is also expensive, such a case could speed-up the application when applied to large source samples since the convolution itself is fast using FFT, giving a run-time scaling faster than that detailed here.

\subsubsection{Advantages and Caveats}

We have motivated a way to produce full posterior distributions of cluster parameters based on the assumption of an underlying mass profile model which can be related to lensing observables, and a priori knowledge of the intrinsic size-magnitude distribution. The main strengths in utilising such a technique lies in the flexibility of the method: complications and extensions can be easily added through explicit marginalisation of latent variables provided they can be related to the observables and intrinsic quantities, and this is done explicitly in the marginalisation over an a priori redshift distribution. In contrast to the frequentist analysis described in section \ref{sec:TheoryAndMethod}, in this formalism any intrinsic correlation between the size and magnitude measures is encompassed in the intrinsic size-magnitude distribution, negating the need for any correction. Further, the method can be applied to produce lens mass profile constraints for each source individually, simplifying the interpretation of the measurements for a chosen sample of sources.

The use of a priori distributions means that the method can be easily implemented using well-motivated theoretical models, or using measurements from the data where available. As such, the application can be entirely self-consistent. However, where the model is measured from data, one must be aware that noise or systematic uncertainties in the measurements can enter the analysis through their effect on the a priori distributions themselves. Where this is the case, only systematic errors in the measured intrinsic quantities which vary in a spatially dependent way will be problematic, as constant offsets across the whole field will cause a identical shift in both the a priori distributions and the sample, provided they are both equally affected. This will therefore not affect the recovered mass profile parameter interpretation. Noise in the measured distributions can be dealt with by smoothing, or fitting a theoretically motivated model to the data. Similarly, the intrinsic distributions should be constructed from a sample which is representative of the de-lensed source sample. This can be done by constructing the distribution across a large area, where the average magnification is unity, or using a sample of field galaxies.

A particular advantage of the use of this method is the fact that posteriors can be constructed individually for each source galaxy, and individually for each foreground lens before further combination. As such, for an analysis which aims to maximise signal--to--noise by stacking lenses, the application of this method allows one to fit the chosen mass profile model to each lens individually and produce model parameter constraints for the lens sample by combining these likelihoods. This therefore avoids the need to fit a mass profile to the stacked measurement, whose shape can be affected by systematics in each individual lens measurement. An example would be in smearing out the profile towards the centre caused by mis-centering on each lens of the stack. With our approach, the uncertain centroid can be taken as a free parameter in the fit for each individual cluster in the sample.

In the application of this method, we choose to work with full recovered model parameter PDFs until the final stage where a maximum-posterior estimator is used to visualise the results in different contexts. Doing so increases the run-time over the case where statistics are formed from frequentist estimators. Especially in the case where multiple latent variables are marginalised over, this can be computationally expensive, however we note that recent work in advanced statistical techniques such as Hierarchical Bayesian Inference \citep[e.g.][]{2015arXiv150507840A,2015ApJ...807...87S} and advanced sampling methods can go some way to reducing the necessary run-time.

\section{The HST STAGES survey}\label{sec:STAGESDataset}

The Space Telescope A901/902 Galaxy Evolution Survey (STAGES), \citep{Gray:2009p2720} utilised the F606W filter of the Advanced Camera for Surveys ({\it ACS}) of the Hubble Space Telescope ({\it HST}) to image a quarter square degree centred on the A901/2 supercluster. The supercluster is made up of four structures at redshift $z = 0.165$, A901a and A901b in the north and A902 and the SW group in the south. In addition, there is a background cluster (CB1) seen in projection with A902 at redshift $z = 0.46$, determined with the application of a 3D lensing analysis in \cite{Taylor:2004p2808}. STAGES images are complemented by optical imaging using COMBO-17 \citep{Wolf:2003p2805} with five broad bands and twelve narrow bands, and which provides high quality photometric redshifts, with the precision $\sigma_z \sim 0.02(1+z)$ for about $\sim 10\%$ of the brightest galaxies ($R_{\rm Vega}<24$) in the STAGES sample. \cite{2004A&A...421..913W} recommends that the limit of $R_{\rm Vega}<24$ is applied in order to keep the photometric redshift error scatter at less than 7\%.  In \cite{2008A&A...480..703H} , an analysis of the COMBO-17 data in the magnitude range $23 < R _{\rm Vega}< 24$, showed that excluding the narrow band data causes the redshift scatter to increase by 30\% and the catastrophic outlier rate to increase by 20\%.  This shows the importance of the narrow-band information in accurate redshift estimation, and also suggests that there would be little gain in using only broad-band information to extend the photometric redshift range beyond $R_{\rm Vega}=24.$  STAGES provides deep ($m_{\rm F606W} \lesssim 27.5$), high resolution HST images of $\sim 70,000$ extended sources, from which a large sample set of robust galaxy shapes, sizes and fluxes can be obtained. The masked observational footprint of the survey covers $\sim 0.22$ square degrees, giving a global number density of sources of $\sim 85 \; {\rm gal/arcminute}^2$ using the whole sample of extended sources. Observations were taken within a small observational time frame, with greater than $50\%$ of the tiles observed in one five-day period, and over $90\%$ within 21 days, whilst seven tiles were observed six months later, minimising temporal (and therefore spatial) variation in the point spread function across the field.  The mosaic of 80 ACS tiles which constitutes the STAGES field is shown in \cite{Heymans:2008p2060}, grouped in colour by observation period.

The application of the analysis outline in section \ref{sec:Reconstruction_Method} to the STAGES data provides a unique set of idiosyncratic complications. The biggest complication is the lack of redshift information for approximately $90\%$ of the source sample. This affects the analysis as presented in two ways: firstly, the lack of multi-band photometry for this sample complicates the removal of cluster members, detailed further in the next section; and secondly, without redshift information we must marginalise over an a priori redshift distribution for the sources to convert lensing observables to cluster mass profile parameters. Following the shear application in \cite{Heymans:2008p2060}, we model the redshift distribution as 
\be\label{eqn:RedshiftDist}
p(z|m_{\rm F606W}) = \frac{3}{2z_0\Gamma ( 2 )}\left(\frac{z}{z_0}\right)^2 e^{-(z/z_0)^{1.5}},
\ee
with $z_0 = z_{\rm median}/1.412$, and using the median-redshift magnitude relation of \cite{Schrabback:2007p2802}
\be\label{eqn:zmed_mag}
z_{\rm median} = 0.29[m_{\rm F606W} - 22] + 0.31.
\ee

\subsection{Mass Profile Modelling}

We model the mass profile of the lensing clusters as spherically symmetric NFW profiles \citep{Navarro:1997p2675}, and relate the profile parameters to lensing parameters by projecting along the line of sight using the analytic relations of \cite{Wright:2000p2260}. The base model profile is a function of four parameters, namely the position of the centre of the profile (centroid), the redshift of the lens, the virial radius/virial mass and the concentration. With the exception of CB1 at $z = 0.46$ \citep{Taylor:2004p2808} all lenses are placed at a fixed redshift $z = 0.165$ \citep{Gray:2009p2720}. Following the shear analysis, we use the mass-concentration relation of \cite{Dolag:2004p2721}, and take the cluster centre positions to be those quoted in \cite{Heymans:2008p2060}. As a result, the NFW fit is a function only of the virial mass/virial radius. Whilst we note that more recent mass-concentration relations exist, and emphasise that the centroid position and concentration could be simultaneously fitted using this method, the over-riding aim of this analysis is to compare cluster profile estimates between the shear and magnification analyses, and so we choose to set up the analysis using the same assumptions as \cite{Heymans:2008p2060} to facilitate comparison.

\subsection{Source Selection}\label{sec:Source_Selection}

We analysis the source catalogue used in the analysis of \cite{Heymans:2008p2060} (hereafter referred to as H08), matched to the publicly available STAGES catalogue \cite{Gray:2009p2720} (hereafter G09). The H08 catalogue consists of $79,366$ sources ($n_{\rm gal} = 96.5$ sources per square arcminute) with SExtractor \citep{1996A&AS..117..393B} MAG\_BEST magnitude information, whilst the G09 source catalogue with a higher detection threshold contains a total of $46,471$ sources ($n_{\rm gal} = 56.5$) with SExtractor and GALFIT \citep{2002AJ....124..266P} size and magnitude measures, as well as COMBO-17 redshift estimation for $10,790$ sources after matching. 

An investigation into size measurement with quadrupole moments (detailed further in Appendix \ref{sec:STAGES_RRG_Measurement}) found that model-fitting methods provide a more accurate size determination for low surface brightness sources in comparison to non-parametric measures such as quadrupole moments or aperture sizes which cannot distinguish between faint, large galaxies and bright, small galaxies. The final source sample therefore consists of the SExtractor aperture magnitude information (MAG\_BEST) as given in the H08 catalogue, chosen to give the largest source sample with magnitude information, with GALFIT half-light radius determined using the GALAPAGOS data pipeline and sky subtraction \cite{2012MNRAS.422..449B} used as the source size where available \citep[see ][ for further details on the source catalogue and the source magnitude and size determination.]{Heymans:2008p2060,Gray:2009p2720}. It is assumed that galactic dust variation across the field is negligible due to the small survey area.

\begin{figure*}
\centering
\includegraphics[width = 0.9\textwidth]{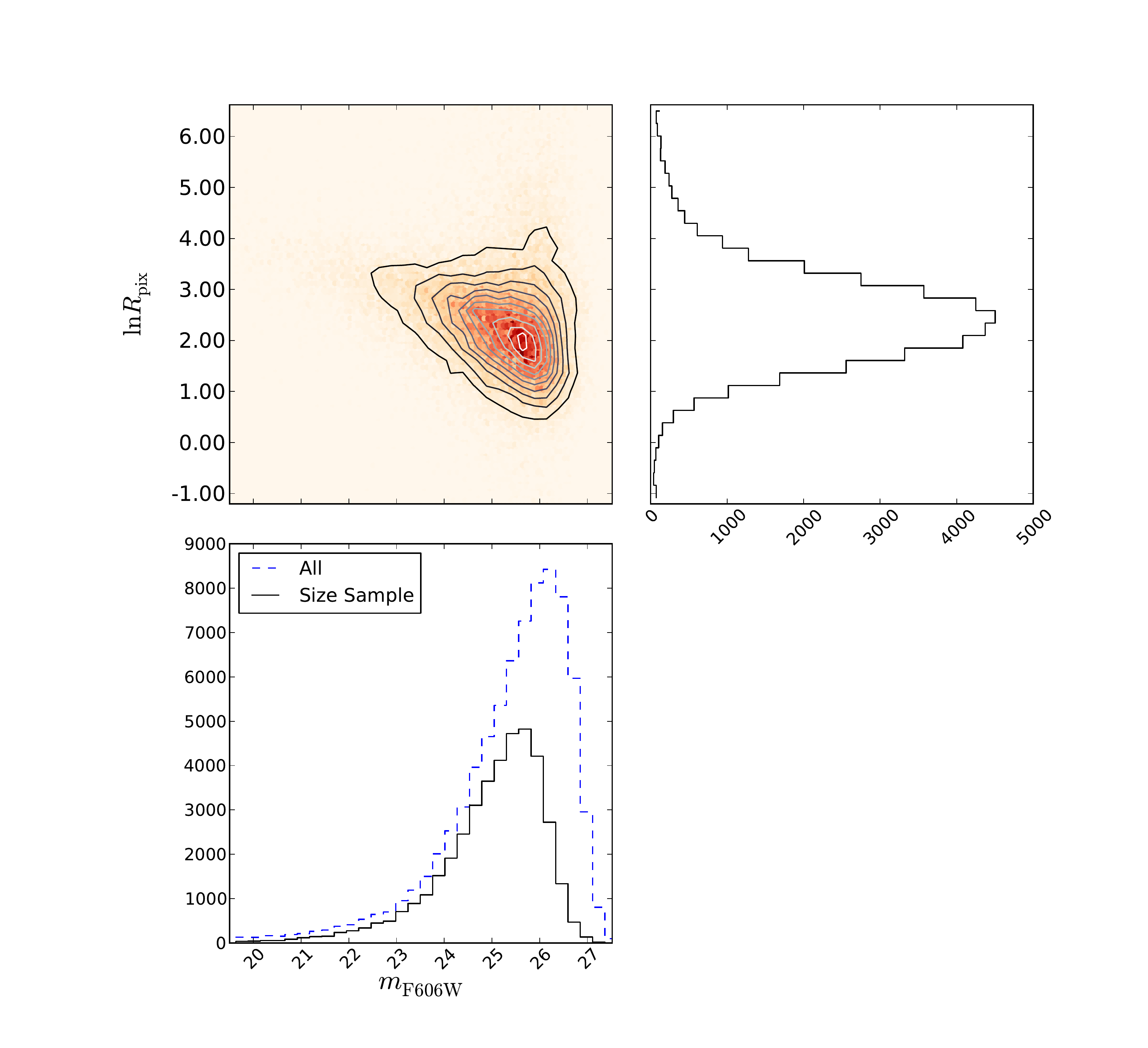}
\caption{The joint size magnitude (upper left) and marginalised size (upper right) and MF606W magnitude (lower) distributions used to for the a priori size-magnitude distribution. Black solid lines show the distributions obtained from the sources in the matched H08 and G09 catalogues, whilst the blue dashed line shows the magnitude distribution for the H08 catalogue only} \label{fig:SizeMag_Distribution_Master}
\end{figure*}

 The a-priori size and magnitude distributions are constructed from the full catalogue after masking conservative $3$ arc-minute apertures around the brightest central galaxy (BCG) cluster centres to remove cluster members. The measured distribution is given in Figure \ref{fig:SizeMag_Distribution_Master}, and forms the a priori distribution for this analysis. The bottom panel of Figure \ref{fig:SizeMag_Distribution_Master} shows the marginalised magnitude distribution between the H08 and G09 catalogues. One can see that the marginalised magnitude distribution for the H08 catalogue extends to fainter magnitudes than the public G09 catalogue, reflecting the different selection of the source sample, where the H08 catalogue includes smaller and fainter sources used in the shear analysis of H08. 
 
In the application of the method, the a priori size-magnitude distribution is constructed and smoothed using Kernel Density Estimation (KDE), using a bivariate-Gaussian smoothing window in size and magnitude, with covariance equal to $0.01$ times the covariance of the data sample. KDE-smoothed apparent magnitude and size distributions constructed in this manner compare well to histograms of the same quantities.

\begin{figure*}
\centering
\includegraphics[width = 0.8\textwidth]{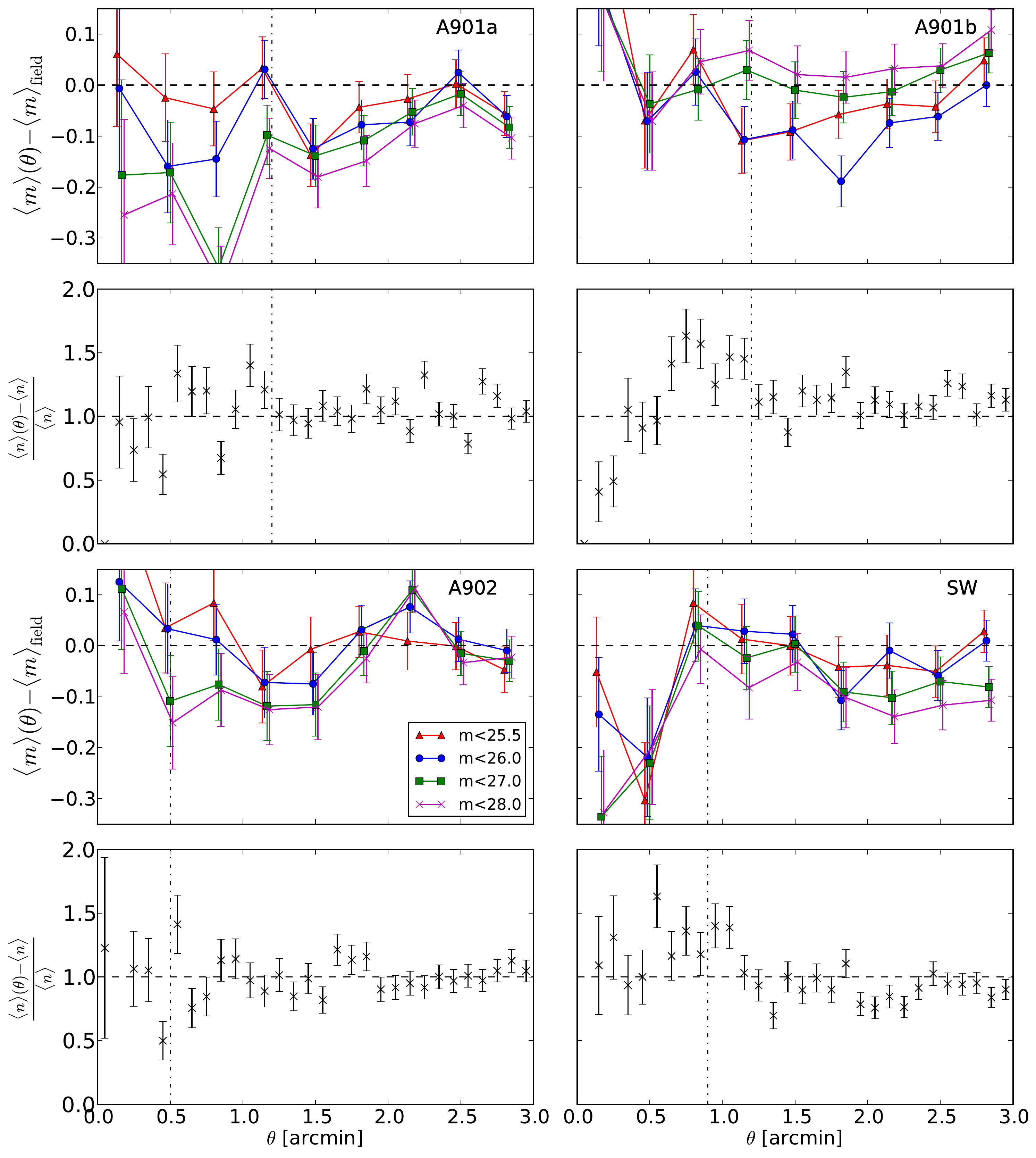}
\caption{Plot showing the difference between the mean magnitude and fractional number over-density of sources in radial bins from the BCG against the field mean, for A901a (top left), A901b (top-right), A902 (bottom-left) and SW (bottom-right), as a function of limiting faint magnitude.} \label{fig:magDiff_byMeasure}
\end{figure*}

The inadvertent inclusion of cluster members in the source sample can introduce a bias in the derived cluster model parameters, as they are mistakenly interpreted as lensed sources in the analysis.  This is a particular problem in the application to the STAGES dataset, as COMBO-17 redshift information is only available for $\sim 10\%$ of the sample, meaning a simple redshift cut is unlikely to remove all cluster members from the sample. We cut source with $z<0.2$ where redshift information is available, and sources brighter than $m=23$, corresponding to a median redshift of $z=0.6$ in the median-redshift-magnitude relation of equation \ref{eqn:zmed_mag}, following H08. The lower panels of Figure \ref{fig:magDiff_byMeasure} show the number density contrast of sources in annular bins around the BCG for each of the four main clusters considered, after the application of redshift and magnitude cuts on the sample. One can see that even after the application of such cuts, the number density of sources is higher than the field average towards the centre of the cluster, most noticeably for A901b and SW, with an amplitude larger than can be accounted for from magnification bias alone.  This suggests that the applied bright magnitude and redshift cuts are insufficient to fully remove cluster members. 

The top panels of figure \ref{fig:magDiff_byMeasure} shows the difference between the mean magnitude in radial bins around the BCG of each cluster to the field mean (after the masking of the four clusters) as a function of varying the faint limiting magnitude of the source sample. Each magnitude difference can be related to the average magnification factor for sources within that annulus, however one must note that this measure has not been corrected for the application of size and magnitude cuts, and is therefore not an unbiased estimate of the cluster mass. However, the use of this estimate can give a useful diagnostic on the behaviour of the signal around the cluster centre.  One can see that, for A901a, A902 and SW the magnitude difference in radial bins is well behaved at large radii, with a general trend towards more negative values as the faint limit used is relaxed. This behaviour may be attributed to the lack of correction for the use of a magnitude cut: where a global faint cut is applied, the mean measured around a magnification field will be underestimated. By contrast, for A901b, the use of a brighter faint cut shows the opposite trend, and we see that for A901b the magnitude difference using the $m<26$  sample is discrepant with more relaxed cuts. This indicates that the signal around A901b is sensitive to the limiting magnitude, and provides a flag to the reliability of the magnitude estimation of the faint sources in that region. We note that A901b shows the largest extended X-ray emission on the STAGES field, and consequently the reliability of the magnitude determination of the faint sources could be compromised by the presence of unaccounted-for intra-cluster light erroneously adding flux to the galaxies behind A901b. As a result, the sources chosen around A901b are taken to be those which satisfy $m<26$ such that the extra intra-cluster light is sub-dominant to the galaxy flux. In this case, the sample of sources around A901b are considered as a separate sample to the remaining sample, and the application of a stricter magnitude cut requires that the posteriors obtained for each of these galaxies must be correctly normalised to account for this.  

Motivated by the trends described here, we therefore apply core cuts on the sample of $1.2', 1.2', 0.5',$ and $0.9'$ around the A901a, A901b, A902 and SW BCGs respectively (shown as dot-dashed vertical lines in Figure \ref{fig:magDiff_byMeasure}). Sources are selected in $3$ arc-minute apertures around the cluster BCG, taken from Table 1 of H08.  In the case of A901b the number over-density of sources extends across the whole angular scale considered here suggesting that cluster member contamination may persevere in spite of the application of source removal within this aperture, however stricter cuts will remove progressively more of the source sample, and will leave only those sources furthest from the cluster centre which are least lensed and whose lensing parameter determination is expected to be most noisy.

The application of the mask around the cluster core provides a natural minimum physical length scale on source-cluster separation, as the removal of the cone around the centre of the cluster means that no source can be closer than the physical distance between the cluster centre and the edge of the masked region on the cluster redshift plane. As well as limiting cluster members, such a cut has the further advantage of reducing the effect of any intrinsic size- or magnitude-density correlations in addition to the redshift and magnitude cuts applied to limit the presence of sources radially close to the lens.

The application of such cuts removes a significant fraction of the sources for which the lensing signal will be strongest, removing $402, 515, 70$ and $282$ galaxies from the source sample around each cluster respectively, with a further $1194$ faint sources removed around A901b after the application of a faint cut of $m<26$. The need to apply such strict core cuts should be considered a particular limitation of the data-set used, and cluster model parameter values would be constrained to higher significance in a data-set with more complete redshift information by allowing the sample to be sufficiently cleaned of sources close to the lens without the application of conservative blanket cuts, or allowing the application of a model to account for the presence of unlensed cluster members or intrinsic size and magnitude-density correlations. Alternatives to the source selection criteria here which avoid the removal of sources from the sample are considered in Appendix \ref{Sec:Alternative-Source-Selection}.

\section{Application to Mocks}\label{sec:ApplicationToMocks}
In this section, the method described in Section \ref{sec:Reconstruction_Method} is applied to mock catalogues, to ascertain the level of statistical error expected of an application of the method to HST data, and to quantify any inherent biases in the analysis. 

\subsection{Mock Catalogue Construction}\label{sec:Mock_Catalogue_Construction}
Mock catalogues are constructed to mimic the STAGES dataset using the following process:

\begin{enumerate}
\item{Galaxies are randomly positioned in the mock survey field.}
\item{Each mock galaxy is assigned an intrinsic magnitude, size and signal--to--noise ratio randomly sampled simultaneously from the STAGES catalogue. This preserves the form of the size and magnitude distributions in the STAGES field, including any size, magnitude and signal--to--noise ratio correlation. We consider two samples here: the ``GALFIT sample'' samples GALFIT sizes and SExtractor magnitudes directly from the Master catalogue, and as such considers the case where a subset of the STAGES sources have valid size measurements, and therefore most closely reflects the application to the STAGES field; the ``All Sizes'' sample samples quadrupole measured sizes (see Appendix \ref{sec:STAGES_RRG_Measurement}) and SExtractor magnitudes from the H08 catalogue, and considers the idealised case where all sources have valid size measurements.}
\item{Each galaxy with $m>23$ is assigned a redshift  randomly sampled from a redshift distribution given by equation \ref{eqn:RedshiftDist} with median redshift given by the median-redshift-magnitude relation of \cite{Schrabback:2007p2802} measured on the GOODS field.}\label{Mock_Construction__Redshift_Assignation}
\item{Unlensed distributions are output, where all redshifts are discarded for the unlensed STAGES mock catalogue, and where a mock ``COMBO'' subset of galaxies is constructed by randomly sampling a sub-set of $10\%$ of the full STAGES mock. The COMBO mocks will therefore vary qualitatively from the observed COMBO-17 sub-sample of STAGES galaxies with redshift information: in the observations, redshifts are obtained only for the brightest galaxies, whilst no magnitude cuts are applied in the construction of the COMBO mock catalogue; as such the mock will have an overall larger median redshift than the observations. The COMBO mock catalogues considered here are constructed with the purpose of testing the sensitivity of the method to the change in number counts and redshift knowledge that results from the application of the method to the sub-set of STAGES galaxies with COMBO-17 redshift information, and are not constructed to be fully representative of that sample.}\label{Mock_Construction__UnlensedOutput}
\item{Each galaxy has its size and magnitude altered according to the lensing relations given in equations \ref{eqn:Lensing_Relations__Size} and \ref{eqn:Lensing_Relations__Magnitude} respectively. The weak lensing limit is therefore not enforced for the magnification relations. Each galaxy is assigned a local magnification due to a set of foreground clusters, modelled as NFW profiles, where the redshift information from the previous step is retained and used to evaluate $\Sigma_{\rm Crit}$ (equation \ref{eqn:Sigma_Crit}) for each galaxy. Each lensing cluster is placed at a redshift of $z_{\rm lens} = 0.165$, which is the measured redshift of the four largest STAGES clusters. Where only a single mock cluster is considered, the cluster is placed with its centre on the BCG of the A901a cluster. No limitations on the size of the magnification factor are enforced. The rare occasional source which lies within the caustic of the cluster, and therefore experience a negative magnification equivalent to a flip in parity, are removed from the sample.}\label{Mock_Construction__Lensing}
\end{enumerate}

Unless otherwise stated, the intrinsic size-magnitude distribution is constructed from the unlensed catalogue, using the full STAGES dataset even when the COMBO redshift subsample is considered, to reduce noise. No size-redshift relation is enforced, however a redshift-magnitude dependence is enforced by sampling source redshift using the median-redshift-magnitude relation of \cite{Schrabback:2007p2802} in point \ref{Mock_Construction__Redshift_Assignation}.

Mock clusters are modelled as spherically symmetric NFW profiles, where the $\Lambda$-CDM mass-concentration relation of \cite{Dolag:2004p2721} is enforced: thus the model assumed for the mass profile parameter recovery is exact in the application to the mock sample.

\subsection{Application of Method}

The application of the method is chosen to match its later use on the STAGES field. Results are shown using a mask of $0.5'$ around each cluster BCG: whilst the use of a core mask is unnecessary for the idealised cases presented here, this masking of the cluster centre is the smallest of the core cuts used in the application to the STAGES field to remove cluster contaminants, and is included here for consistency. No cuts are imposed on source size, nor on faint magnitudes. Source sizes and magnitudes have negligible measurement error, and as such the method detailed in section \ref{sec:Reconstruction_Method} can be applied exactly.  This application therefore constitutes an idealised case, and one must note that the application of size cuts, PSF confusion or measurement error may cause a decrease in the constraining power of the analysis. 

Posteriors are evaluated on the virial radius by default, and posteriors on the virial mass determined from these results using conservation of probability:
\be\label{eqn:Mass_Radius_Prior_equiv}
p(M_{200}) \propto \frac{p(r_{200})}{r_{200}^2} \propto \frac{p(r_{200})}{M_{200}^{\frac{2}{3}}},
\ee
where $M_{200} \propto r_{200}^3$ was assumed. As a result, even where mass constraints are presented, a flat prior on virial radius has been assumed: this translates to a prior on virial mass which down-weights large clusters. An extension to this analysis could evaluate virial mass posteriors using a prior motivated from a halo mass function, which would also down-weight large clusters. Error bars are calculated as the region above a common posterior threshold which includes $68\%$ of the probability on either side of the mode of the posterior on $\alpha$, assuming a uniform prior for $r_{200} \ge 0$. 

For simplicity a single cluster is modelled on the field, and the posterior distribution can therefore be sampled directly over a grid. In the application to STAGES data, the application of the analysis will utilise an MCMC algorithm to sample the multi-dimensional likelihood parameter space that results from the need to simultaneously fit masses and centroid positions for multiple clusters on a single field. The application on this MCMC algorithm has been compared to mock realisations used as part of this analysis, and has been verified to return the same posteriors as the simplified case presented here. 

\begin{figure}
\centering
\includegraphics[width = 0.5\textwidth]{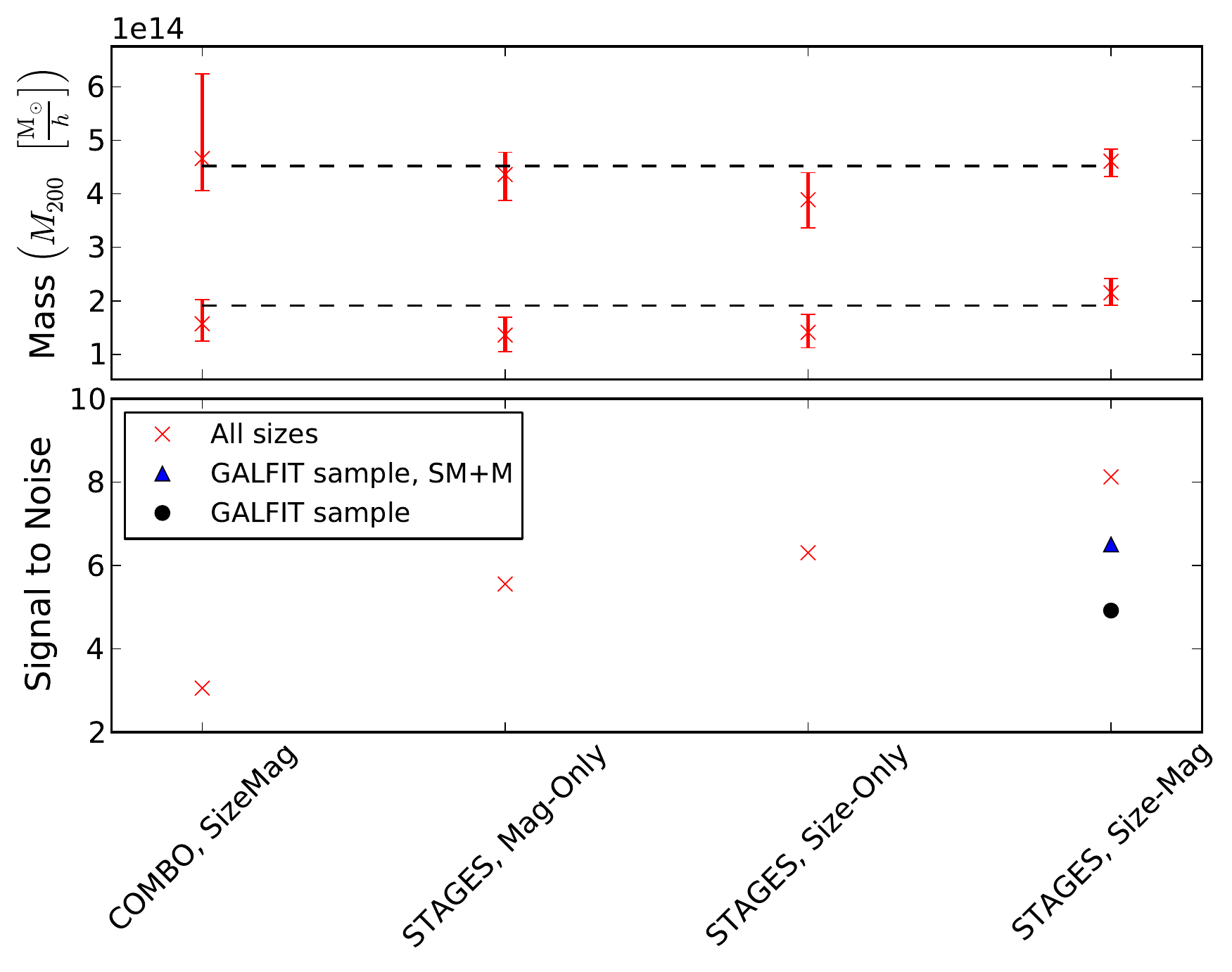}
\caption[Example results for a size-only, magnitude-only and joint size-magnitude cluster reconstruction using mock STAGES and COMBO data, with signal--to--noise]{Comparison plot between the size-only, magnitude-only and size-magnitude analyses for both mock COMBO- and the mock STAGES-datasets. The {\it top} panel shows example analyses for each case. In all cases, a single cluster was modelled on the field to avoid bias due to overlap between clusters, and the prior was constructed on the unlensed STAGES dataset. Errors are 68\% confidence limits of the recovered posterior about the mode position. Dashed lines show the input mass for each case. The {\it bottom} panel shows signal-to-noise, calculated as the mode point divided by half the total error width for each comparison. One can see that the size-magnitude analysis with the full STAGES set gives the largest signal-to-noise of all four cases, motivating its use on the full STAGES dataset. In the latter case, we show the average signal--to--noise--ratio for the case when the source sample reflects only the sub-sample with valid GALFIT size estimates (labelled `GALFIT sample), and when the magnitude information of the sources without sizes are combined with the size-magnitude analysis of the sources with sizes (labelled `SM+M').} \label{fig:Experiment_Comparison}
\end{figure}

Figure \ref{fig:Experiment_Comparison} shows a comparison plot for four different analyses using mock STAGES data-sets, where single NFW clusters have been modelled. The plot considers four different data-sets for the analysis:
\begin{enumerate}
\item{COMBO, Size-Mag: Posteriors are constructed using information on both galaxy size and magnitude. The data-set is limited to only those galaxies with redshift information.}
\item{STAGES, Size-Mag: As above, using the full STAGES data-set, with redshifts for $\sim 10\%$ of sources. Where no galaxy redshift information is present, the likelihood is constructed by marginalising over a redshift distribution. We also consider the case where only a fraction of sources have valid size information (as in the STAGES data, and labelled the `GALFIT sample') giving a reduction in the total source count, and with the addition of magnitude information for these sources (labelled `SM+M').}
\item{STAGES, Mag-Only: As above, using only magnitude information: galaxy size information is marginalised, as detailed in section \ref{sec: Size_Magnitude_Only_method}.}
\item{STAGES, Size-Only: As STAGES, Size-Mag, using galaxy sizes only: magnitude information is marginalised, as detailed in section \ref{sec: Size_Magnitude_Only_method}.}
\end{enumerate}
The top panel of Figure \ref{fig:Experiment_Comparison} shows example single runs for each analysis type for two different input masses. Whilst there is some expected statistical variation between runs, in all cases the input mass is well reproduced.  The bottom panel shows an estimate of the signal--to--noise, constructed as the mode point of the recovered mass posterior for the largest input mass divided by half the total error width, averaged over 10 independent realisations for each data-set considered.  From this one can see three main features: Firstly, one sees a significant increase in the signal--to--noise when the full STAGES data set is used, rather than the sub-set of sources with COMBO redshift information. This is a result of the decrease in statistical noise as a consequence of the increase by a factor of $\sim 10$ in the number density of sources in the full STAGES dataset. Secondly, we find that using the STAGES data-set, there is a small increase in signal--to--noise as one moves from a magnitude- to size-only analysis, with a significant increase when both are used. This latter point is expected as both the size and magnitude information contain complementary information on the magnification field which results from the presence of the lensing cluster. Finally, we note that the `GALFIT sample' gives a much reduced signal--to--noise compared to the full sample: this results from the reduction in the total number density of sources when the number of sources with size information is chosen to most accurately reflect the application to STAGES data, and one can see that the further addition of magnitude information for those sources without size information recovers much of the lost information. These results are summarised in Table \ref{Table:Experiment_Average_Noise}, which shows the average uncertainty in each probe over these mock realisations, for a $2\times 10^{14} h^{-1}M_{\odot}$ cluster similar to A901a or A901b.

\begin{table}
\begin{center}
\caption{The average width of $1\sigma$ error bars taken over 10 mock realisations, for each probe considered in Figure \ref{fig:Experiment_Comparison}.}\label{Table:Experiment_Average_Noise}
\begin{tabular}{|l|c|c|}
\hline
Input: & \multicolumn{2}{c|}{$r_{200} = 1.2 h^{-1}{\rm Mpc} ;\; M_{200} \sim 20\times10^{13} h^{-1}M_{\odot} $} \\
\hline
\\
Experiment & $\bar{\sigma}_{r_{200}} [h^{-1}{\rm Mpc}]$ & $\bar{\sigma}_{M_{200}} [10^{13} h^{-1}M_{\odot}] $ \\
\hline
COMBO Size-Mag &0.14 & 6.6 \\
\hline
STAGES Mag-Only &0.07 & 3.5 \\
\hline
STAGES Size-Only & 0.07 &  3.1\\
\hline
STAGES Size-Mag & 0.05 & 2.4\\
\hline
\end{tabular}
\end{center}
\end{table}

\begin{figure}
\centering
\includegraphics[width = 0.5\textwidth]{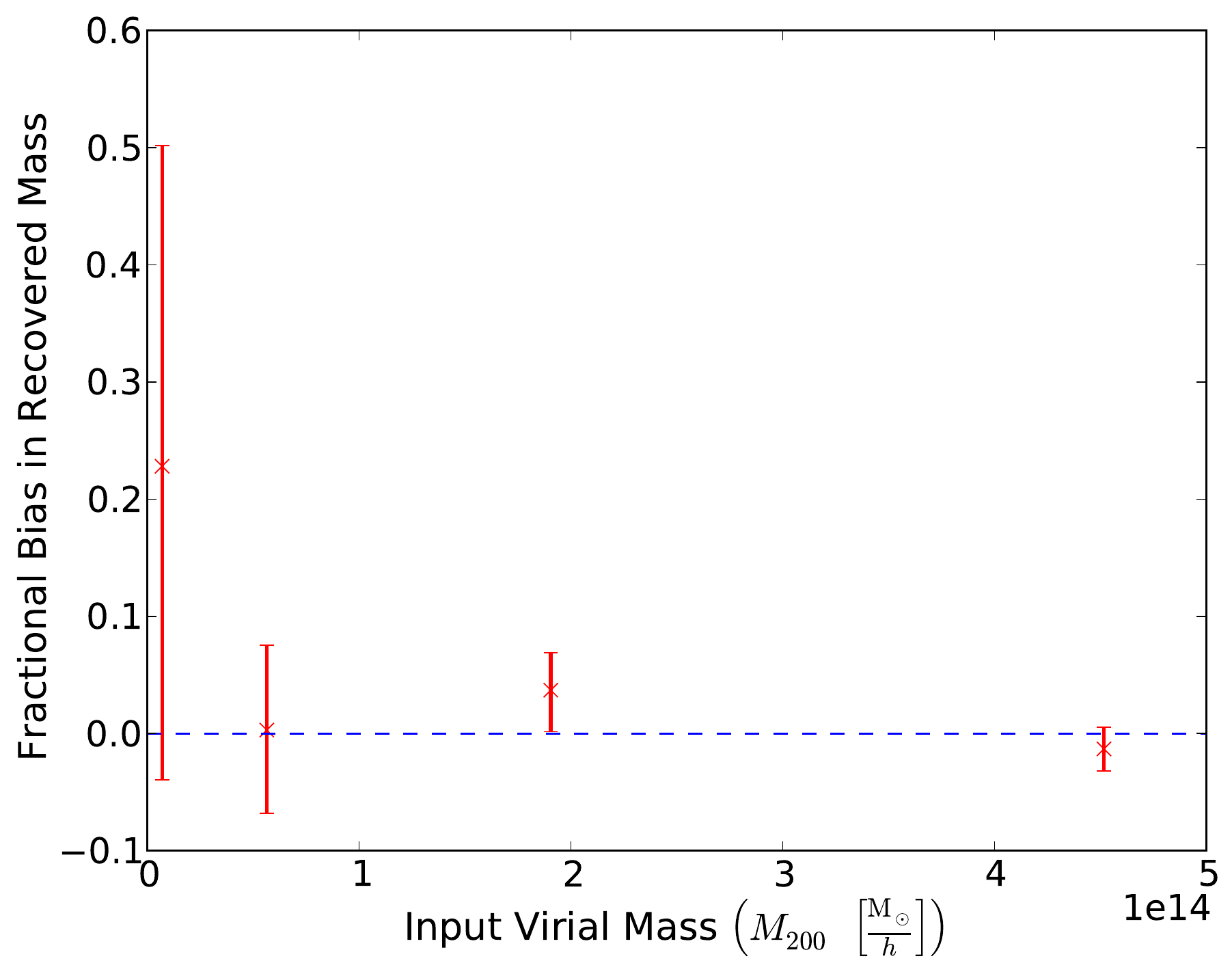}
\caption{Plot showing fractional bias in cluster mass in the application of the joint size-magnification analysis on the STAGES dataset for four input cluster masses. No significant bias in recovered halo mass is evident for the ranges of masses considered here.} \label{fig:Analysis_Bias_STAGES_SizeMag}
\end{figure}

Figure \ref{fig:Analysis_Bias_STAGES_SizeMag} shows the fractional bias, given as
\be
f = \frac{M^{\rm ML}_{200}-M^{\rm Input}_{200}}{M^{\rm Input}_{200}},
\ee
where $M^{\rm Input}_{200}$ is the input mass and $M^{ML}_{200}$ is the mode point of the combined posterior across 10 mock catalogue realisations, and error bars give the 68\% confidence interval on either side of the mode-point of the combined posterior. We see no evidence for significant bias in the application of the method to the simplified case presented here. Posterior construction using the size-magnification or flux-magnification effects individually have also been verified to be similarly unbiased, but as the expected signal--to--noise is largest for the joint analysis, we will focus on the use of the joint size-magnitude analysis on the full STAGES data-set for the remainder.

The results presented here suggest that the use of the joint size and flux magnification signals can probe the large clusters of the STAGES field to high significance for the idealised case considered up to this point. Stricter core cuts such as those motivated for the data will reduce the number density of the sample and result in an increase in statistical noise and subsequent reduction in signal--to--noise ratio over those presented here, where cluster contamination has been largely neglected. 

The presence of cluster members in the source sample, and non-negligible measurement error on source size and magnitude may also introduce bias in the recovered posteriors. We consider the effect of such contamination on the accuracy of the method in the next two sections.

\subsubsection{Bias due to cluster member contamination}

In section \ref{sec:Source_Selection}, we noted that the presence of cluster members in the source sample can introduce a bias in the recovered cluster when not accounted for in the a priori redshift distribution of the sources. In that section, we detailed the methods by which the source sample was selected to minimise this effect, including the application of core cuts around the main over-densities in the field, and the use of a bright magnitude cut as well as a redshift cut where the source also falls into the COMBO-17 sample. In this section, we consider the effect of cluster contamination on the recovered mass for the STAGES clusters after the application of such cuts.

Cluster contamination is modelled in the mock catalogues by constructing a cluster member catalogue according to the cluster contamination profile shown in Figure \ref{fig:magDiff_byMeasure}, where each annulus bin is assigned a number of cluster contaminants given by 
\be
N_{\rm Contaminant} = fn^{\rm mock}_{\rm global}\Omega - N_{\rm annulus}^{\rm Poisson},
\ee
where $f = n_{\rm annulus}/n_{\rm global}^{\rm data}$ as measured from Figure \ref{fig:magDiff_byMeasure}, $n^{\rm mock}_{\rm global}$ is the global number density of sources in the un-contaminated mock catalogue, $\Omega$ labels the area of the annulus, and where $N_{\rm annulus}^{\rm Poisson}$ is the number of sources in that annulus in the un-contaminated mock catalogue. Where $f\le 0$, no cluster members are added to the catalogue. Thus, the cluster catalogue is constructed such that the contamination fraction of the mock is equal to that measured in the data where $f\ge 0$. The cluster members are randomly placed within the annulus, with a size and magnitude jointly randomly sampled from the reference data catalogue with $m\ge 23$ (to mimic the data cuts used in the construction of the contamination profile of Figure \ref{fig:magDiff_byMeasure}), and assigned a redshift of $z = 0.165$.  This cluster catalogue of unlensed members is concatenated with the original source catalogue after the source catalogue has been lensed by a model NFW profile. Each cluster is modelled individually, and a core aperture mask of $1.2'$ around A901a, $1.2'$ around A901b, $0.5'$ around A902 and $0.9'$ around SW is applied in the application of the cluster mass measurement to mimic the application to data. 

\begin{figure}
\centering
  \begin{tabular}{@{}c@{}}
    \includegraphics[width=0.5\textwidth]{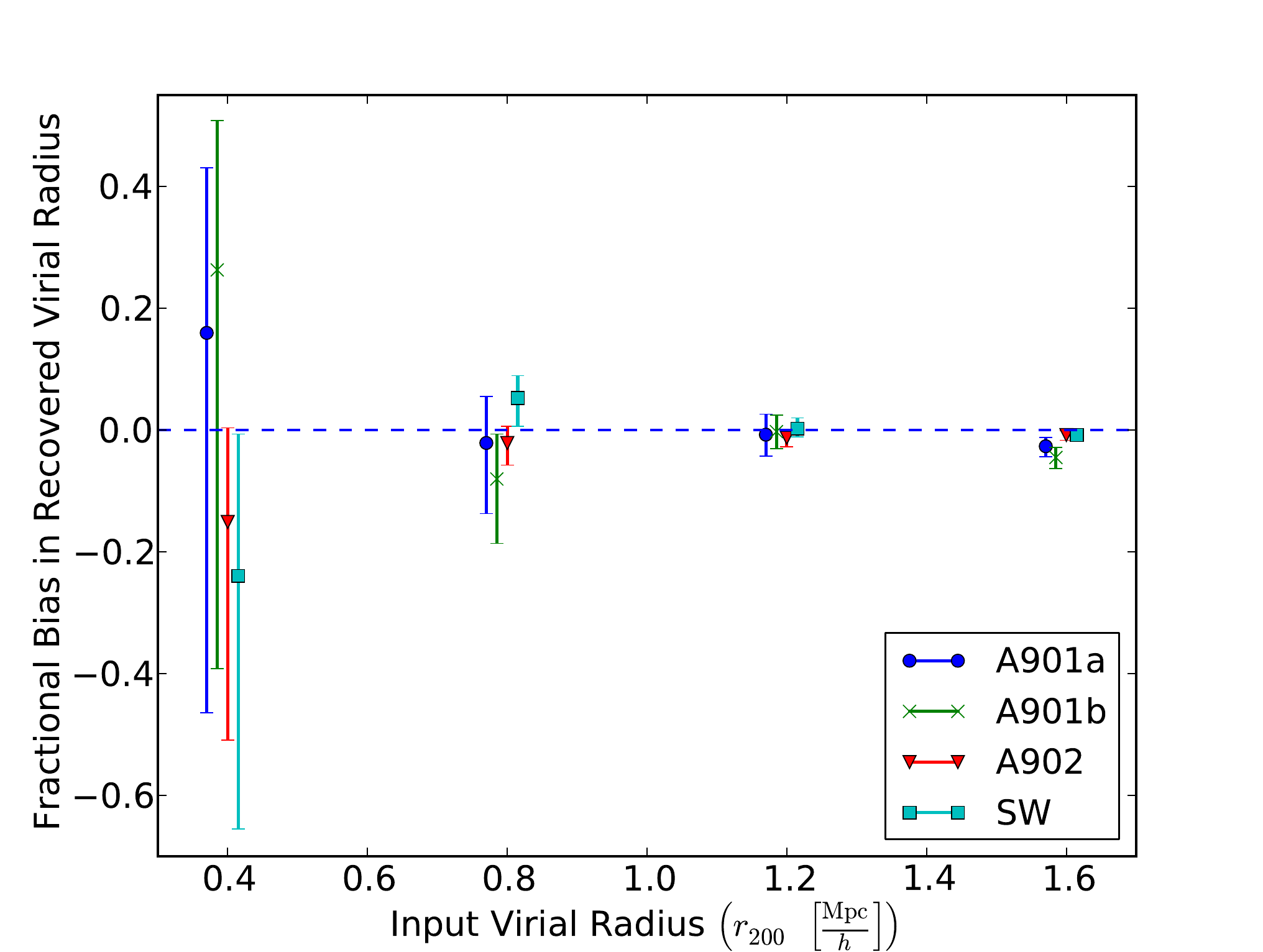} \\
  \end{tabular}
  \caption[Fractional bias on recovered cluster virial radius due to a sample of unremoved cluster contaminants]{Figure showing the effect of a sample of cluster contaminants on the recovered posterior on the cluster virial radius, showing the fractional bias in virial radius for the contaminated catalogue as a function on modelled virial radius. Data points are slightly offset in the x-value to aid visualisation. All values are calculated over 10 mock realisations of the catalogue, where each cluster is modelled individually to avoid overlap bias. The posterior is calculated for each cluster using the core masking for that cluster detailed in section \ref{sec:Source_Selection}, with number of contaminant clusters chosen to match the profile of Figure \ref{fig:magDiff_byMeasure} where an over-density is observed.}\label{fig:ClusterContamination_Mocks_SizeMag}
\end{figure}

Figure \ref{fig:ClusterContamination_Mocks_SizeMag} shows the average signal--to--noise and fractional bias on the recovered virial radius using a joint size-magnitude analysis over 10 mock realisations using the above method of mock construction. In each case, the cluster is modelled individually to avoid overlap bias, and is positioned on the measured BCG of A901a. Cluster contaminants are added in annuli up to $3'$ from the centre of the cluster. Typically, A901a contains 140 contaminants ($\sim 7\%$ of total sources within $3'$), A901b contains $\sim 300$ ($\sim 16\%$), A902  $\sim 60$ ($\sim 3\%$) and SW $\sim 80$ ($\sim 4.5\%$) when core masking is not used.  The top panel shows the fractional bias for each modelled cluster as a function of input cluster virial radius. We see that in the presence of cluster contaminants, there is no strong evidence for bias amongst all modelled clusters, with the possible exception of A901b which shows evidence of a small negative bias of a few percent, particularly at larger virial radius where the statistical noise is smallest.

Figure \ref{fig:ClusterContamination_Mocks_SizeMag} indicates that the choice of core masking aperture applied to the data is sufficient to remove any bias caused by cluster contamination of the sort considered here.  However, one must note that these results consider a particular simplified form of cluster contamination, with only the inclusion of an unlensed contaminant sample, and does not account for intrinsic magnitude- or size-density correlations due to physical processes during galaxy formation. The scale of such correlations is subject to current investigation, with seemingly contradictory results presented using a variety of surveys and source selection methods complicating the choice of an appropriate model \citep[see ][ for a short review of recent measurements.]{Alsing:2014p2846}. The investigation into the impact of these effects is therefore considered outwith the scope of this investigation, and left to future work.

\subsubsection{Bias due to measurement noise}\label{sec:NoiseBias}

In Section \ref{sec:Reconstruction_Method}, we noted that the pipeline as detailed does not explicitly account for measurement error on measured source size and briefly detailed how one may edit the likelihood evaluation to account for measurement noise in any of the observed quantities. In this section, we quantify the expected bias due to unaccounted-for error in the measured size and magnitude in the idealised STAGES catalogue.

In the method of mock catalogue construction detailed in Section \ref{sec:Mock_Catalogue_Construction}, it is assumed that the measured sizes are exact, and that any variation in measured size is due only to lensing by foreground structure. Measurement noise is included in the mock catalogue construction by adding an uncertainty sampled from a Gaussian distribution with width $\sigma_R = 0.2R$ and $\sigma_m = 0.08$, after sizes and magnitudes have been sampled from the master catalogue and before lensing by the simulated cluster. The size uncertainty used approximately corresponds to the measured uncertainty in PSF-corrected quadrupole sizes for the high signal--to--noise sources considered in Appendix \ref{sec:STAGES_RRG_Measurement}, and the average measured uncertainty in the GALFIT scale radius in bins of measured scale radius taken directly from the G09 catalogue. The magnitude uncertainty is taken from the mean MAG\_BEST uncertainty across the entire field. The measurement noise is included in the unlensed catalogue, which is used to construct the a priori size distributions for the application to mocks, as well as the source sample in the measurement with the pipeline. 

\begin{figure}
\centering
\includegraphics[width = 0.5\textwidth]{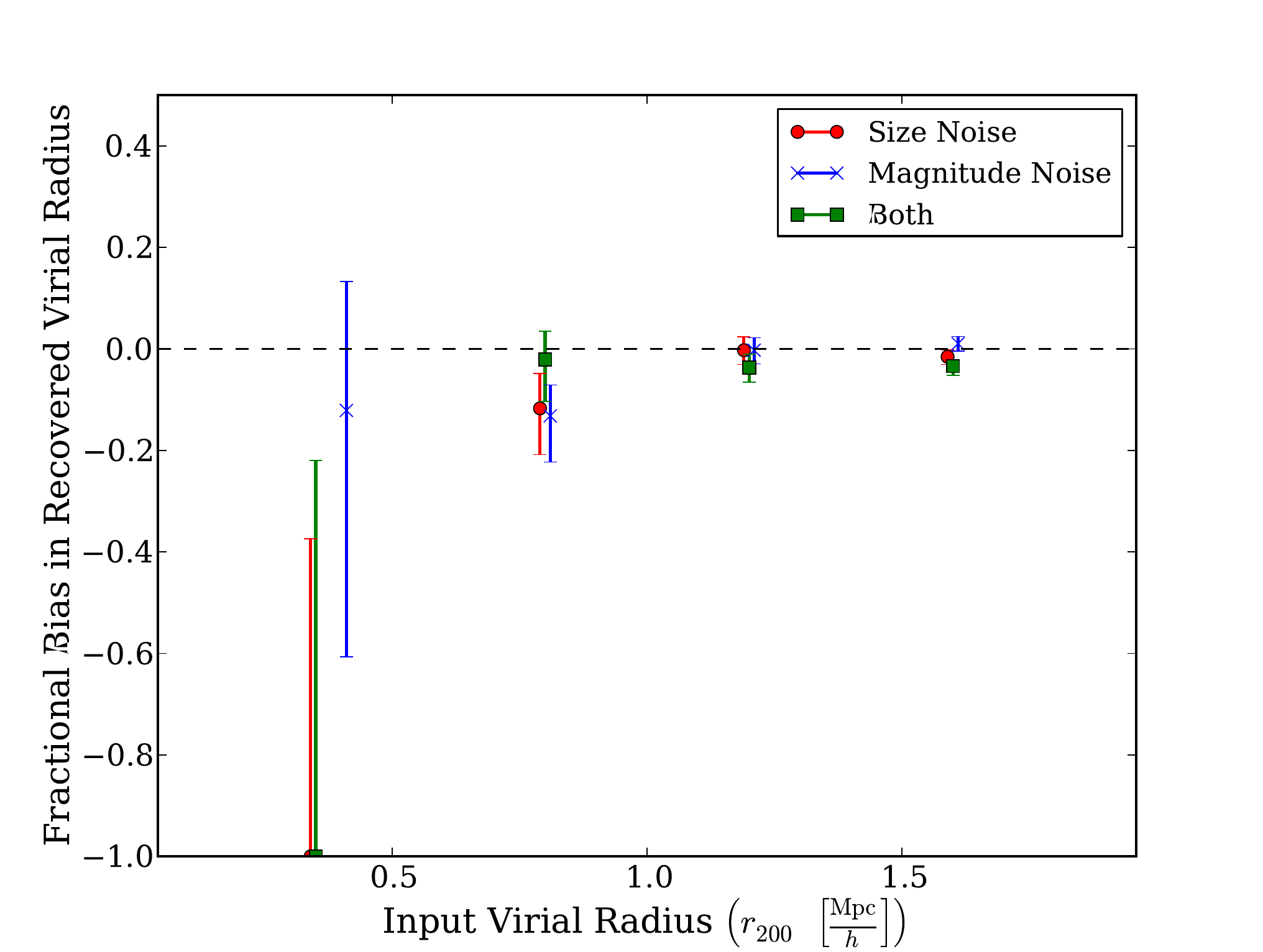}
\caption{Plot detailing the bias in recovered virial radius from a joint size-magnitude analysis resulting from noise in the size and magnitude measurements, when this is not taken into account in the analysis. Red circles correspond to the application of Gaussian noise on size with mean zero and  $\sigma_R = 0.2R$, blue crosses a constant Gaussian noise on magnitude with with $\sigma_m = 0.08$, and green squares the combination of both. Co-ordinate values are offset for ease of visualisation, and each group of three corresponds to simulated NFW clusters with virial radii of $r_{200} = 0.4,0.8,1.2,1.6$ $h^{-1}$Mpc respectively. The fractional bias for the `size' and `both' cases for $r_{200} = 0.4$ $h^{-1}$Mpc correspond to $-1$, equivalent to the maximum-posterior point of the combined posterior occurring at $r_{200} = 0.$}\label{fig:Measurement_Noise_Bias_SM}
\end{figure}

Figure \ref{fig:Measurement_Noise_Bias_SM} shows the fractional bias in recovered virial radius over ten mock realisations, for four input cluster masses. We see that there is evidence for a negative bias in the recovered radius, whose absolute value decreases with increasing cluster mass, corresponding to decreasing bias with increasing signal--to--noise. In practice, this would suggest that the measurement of cluster mass for A902 and the SW group should be more affected by noise bias than the larger clusters, with a predicted $\sim 10\%$  bias in virial radius. This bias is smaller than but comparable to the expected uncertainty on the recovered radius for each of these clusters. In practice, one may take the uncertainty on the measured size and magnitude into account using the method presented in Section \ref{sec:Reconstruction_Method} by marginalising over a latent variable which describes the distribution of the measured size or magnitude around the true underlying value. 

Section \ref{Sec:Includ_Measurement_Noise} describes how the method as applied can be extended to naturally account for uncertainty in the size, magnitude and redshift which may reduce the level of bias in cluster parameters predicted here, with a consequent increase in run-time that can easily cause the analysis to take a prohibitively long time to complete without the use of advanced techniques, and is therefore left to future projects. We note however that where the source sample is complete in redshift, or where the source sample can be smaller (for example in the application to smaller or more isolated lenses which do not require simultaneous fitting, such as in galaxy-galaxy lensing) the increased run-time may be less limiting, and can therefore instead be considered idiosyncratic to the application on large, spatially close clusters with low source redshift completeness considered here.

\section{Application to STAGES}\label{sec:ApplicationToSTAGES}

In section \ref{sec:ApplicationToMocks}, we have shown that the application of the proposed method of cluster model parameter determination detailed in Section \ref{sec:Reconstruction_Method} provides a means to accurately measure the mass of mock clusters with a STAGES-like data-set, and quantified any biases resulting from simplifications in the pipeline, or limitations in the data. In this section, we apply the method to the STAGES data-sets detailed in section \ref{sec:Source_Selection}. We have quantified cluster model parameter constraints for the STAGES clusters, with a comparison to existing measurements using shear estimates in H08. Following H08, we consider a fit using four clusters (A901a, A901b, A902 and SW), and a 7-cluster fit (where NFW models are placed on A901b, A901a and the infalling X-ray group A901$\alpha$, A902 and the background cluster CB1, and the SW group are split into two component clusters named SWa and SWb motivated by peaks in the shear parameter-free mass reconstruction).

The source sample is split into two independent samples: the first contains those sources for which a reliable source size is available and is used for a full joint size-magnitude analysis; the second corresponds to those sources for which size information is either unavailable or considered unreliable and is therefore considered only as part of an analysis using measurements of source magnitude only. In all cases a lower size cut of $\ln R = 0.78$ is used to remove the smallest sources for which the correction of the PSF is least robust, and are considered only as part of the second sample. This corresponds to a cut of $R < 2.2$ pixels ($\approx 0.11''$), which is equivalent to the cut used in \cite{Schmidt:2012p1106}. The final result is presented as the combination of independent analysis of both samples.

After source selection as detailed in Section \ref{sec:Source_Selection}, the source sample consists of $7966$ sources, with $2189, 1230, 2437$ and $2110$ galaxies around A901a, A901b, A902 and the SW group respectively, for 4-cluster case. Of these, 4288 are used as part of a size-magnitude analysis, whilst 3678 are used in a magnitude-only analysis. For the seven cluster case, the source sample consists of $2189, 2102, 1230, 2437, 2232, 2110$ and $2043$ sources in $3'$ apertures around A901a, A901$\alpha$, A901b, A902, CB1, SWa and SWb respectively, giving a total source sample of 10,112 sources after subtraction of doubly-counted sources. Of this total 5408 are used for a size-magnitude analysis, and 4704 are used in a magnitude-only analysis.
 
\subsection{Mass Reconstruction of the STAGES clusters}

Unless otherwise stated, the cluster virial radii are allowed to vary independently for considered clusters. Constraints are produced using a Metropolis Hastings Markov-Chain-Monte-Carlo method, and convergence of the recovered posteriors is verified by requiring that the marginalised posteriors for each free parameter satisfy $R<1.03$, where $R$ is the Gelman-Rubin statistic \cite{GEL.RUB}.

\begin{figure*}
\centering
\includegraphics[width = 0.95\textwidth]{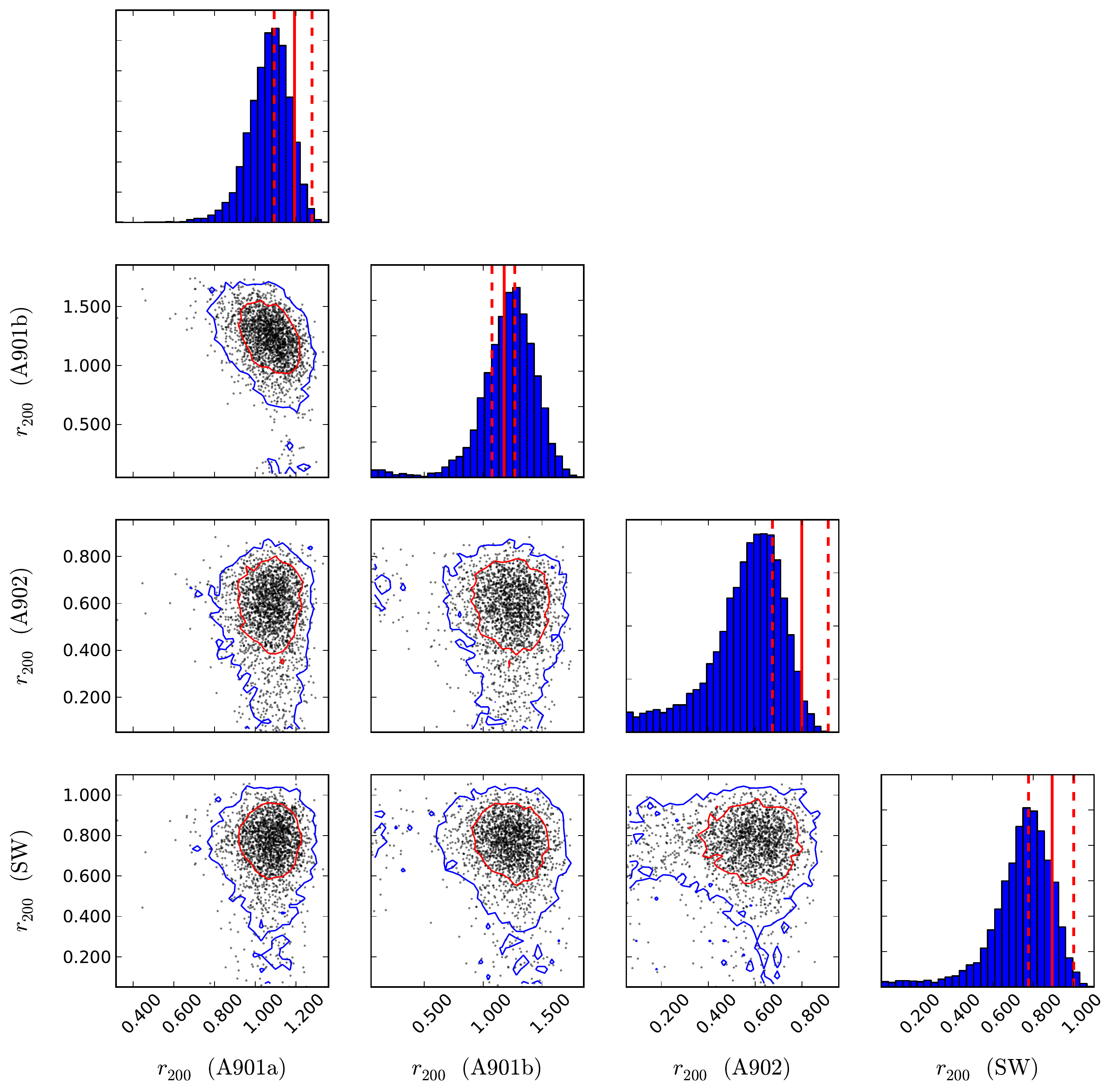}
\caption{Weak lensing magnification constraints on the virial radius for the A901/902 super cluster modelled as four structures (A901a,A901b, A902, SW) centered on their BCGs. Diagonal plots show the marginalised distribution for the virial radius on each cluster, and the vertical red lines show the mean (solid) and 1-$\sigma$ uncertainty (dashed) for the shear analysis given in H08. Off diagonal plots show points from a thinned MCMC chain, and blue and red lines show the 95\%, and 68\% confidence regions for the 2D marginalised distributions respectively.} \label{fig:MassRecon_BCG_4Cluster}
\end{figure*}

Figure \ref{fig:MassRecon_BCG_4Cluster} shows the result where four NFWs are fitted, centered on the BCGs. Diagonal panels show the one-dimensional marginalised posteriors for each single model parameter for each cluster, whilst off-diagonal panels show the two-dimensional marginalised posteriors between two model parameters, with all other parameters across all clusters marginalised over. Vertical lines show the quoted mean (solid) and 1-$\sigma$ uncertainty for the shear analysis of H08. We immediately see that the magnification measurement detects all four clusters, with a signal--to--noise--ratio on the virial radius of $9.3,5.4,3.5,5.1$ for A901a, A901b, A902 and SW respectively.

\begin{figure*}
\centering
\includegraphics[width = 0.95\textwidth]{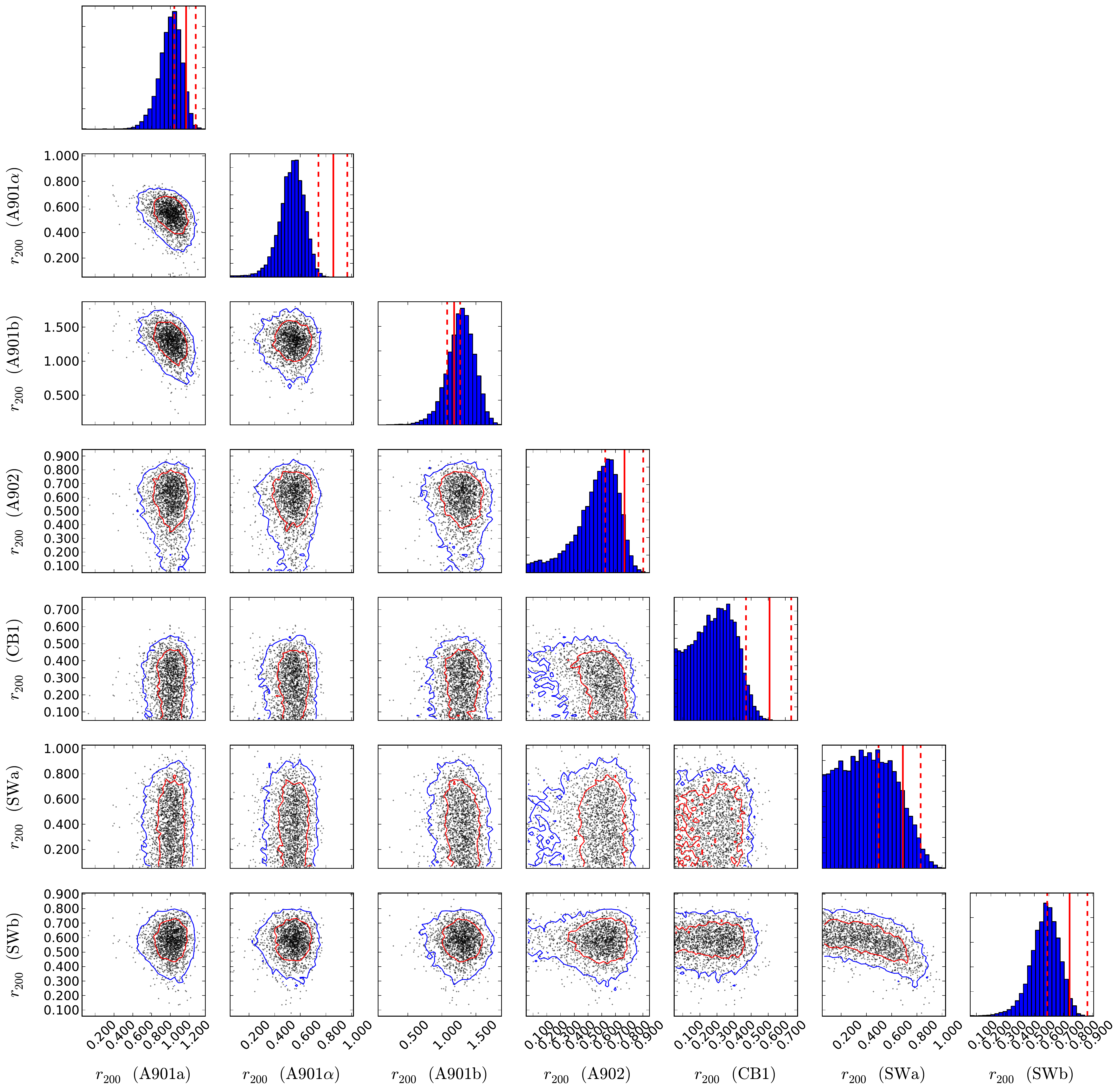}
\caption{Weak lensing magnification constraints on the virial radius for the A901/902 super cluster modelled as seven structures (A901a, A901$\alpha$, A901b,  A902, CB1, SWa, SWb) centered on their BCGs. Diagonal plots show the marginalised distribution for the virial radius on each cluster, and the vertical red lines show the mean (solid) and 1-$\sigma$ uncertainty (dashed) for the shear analysis given in H08. Off diagonal plots show points from a thinned MCMC chain, and blue and red lines show the 95\%, and 68$\%$ confidence regions for the 2D marginalised distributions respectively.} \label{fig:MassRecon_BCG_7Cluster}
\end{figure*}

Figure \ref{fig:MassRecon_BCG_7Cluster} shows the result in the 7-cluster case. In this case, the seven clusters are detected to a signal--to--noise of $7.3, 5.1, 5.4, 3.5, 5.3$ in virial radius for A901a, A901$\alpha$, A901b, A902 and SWb respectively. CB1 and SWb show a maximum-posterior point which is consistent with the presence of a cluster, but with a reduced significance in comparison to the other groups.

\begin{figure}
\centering
\includegraphics[width = 0.5\textwidth]{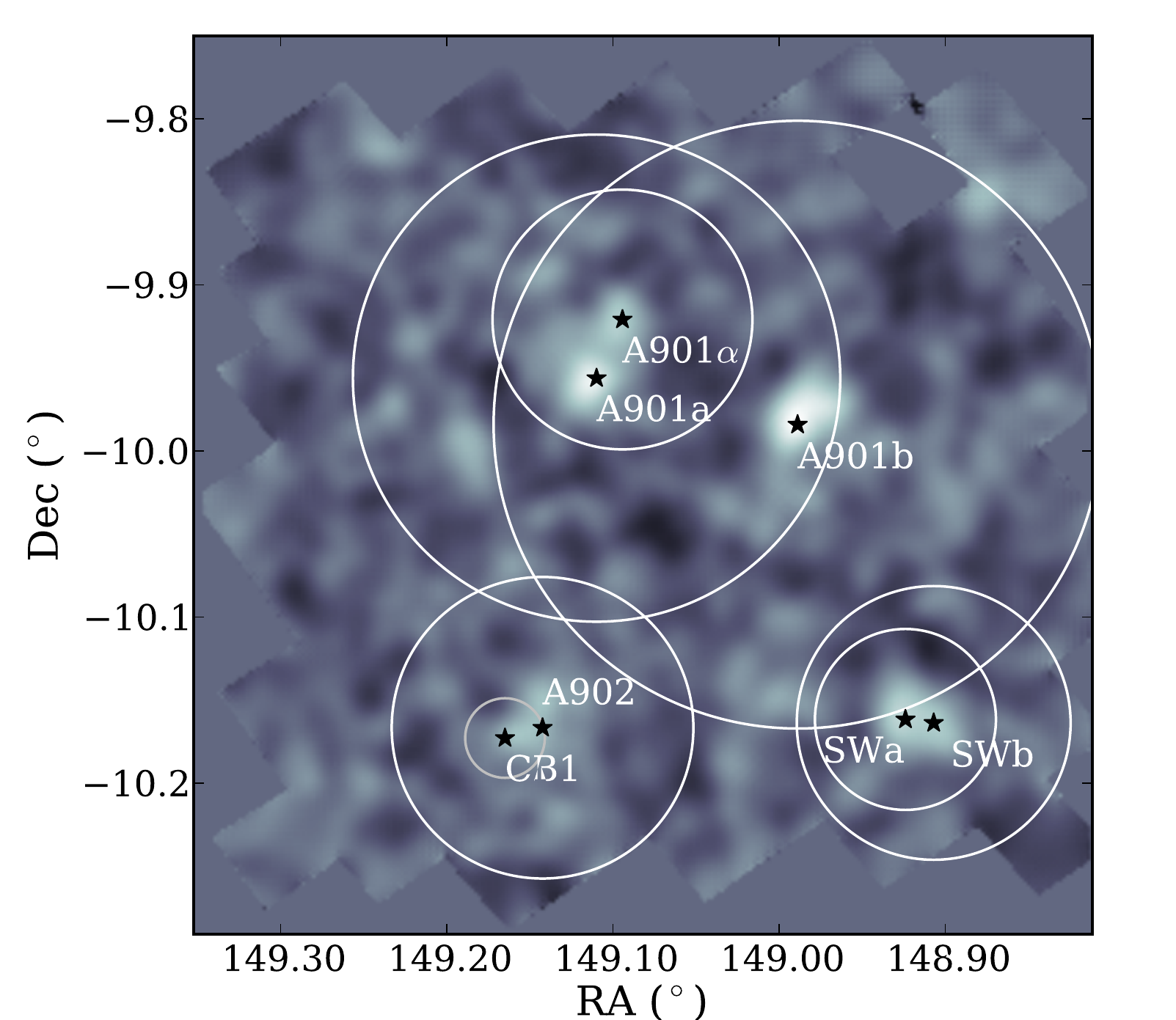}
\caption{Figure showing the cluster centre (shown as stars) and recovered virial radius from the magnification analysis (circles), superimposed over the shear mass reconstruction signal--to--noise plot of H08. The circle denoting the virial radius of CB1 is colored off-white to indicate that the structure exists at a higher redshift to the other structures on the field.} \label{fig:7ClusterField_VirialRadius}
\end{figure}

These results are summarised in Table \ref{tab:NFW_res}, including mass estimates for each cluster considered, and Figure \ref{fig:7ClusterField_VirialRadius} shows the virial radius of each cluster in the `7-cluster' case superimposed on the shear mass reconstruction signal--to--noise map of H08. 

In contrast to the application to the mock catalogues which assumed a flat prior on virial radius in all cases considered, mass estimates here are presented assuming a flat prior on the mass. As a flat prior on the virial radius corresponds to a prior on the mass which diverges as the recovered mass tends to zero (see equation \ref{eqn:Mass_Radius_Prior_equiv}), where the recovered posterior does not tend to zero faster than $M^\frac{2}{3}$ the data is not strong enough to overcome the prior and giving prior-dominated posteriors peaking at $M=0$. We see that this is the case for A902 and SW in the 4-cluster case, and A902, CB1 and SWa in the 7-cluster case, and note that improved data may avoid this issue in future applications. However, the application of a flat prior on mass in this case also allows the direct comparison between these results and those of H08.

\begin{table*}
\begin{center}
\begin{tabular}{l|l|c|c|c|r}
\hline
Structure & RA & Dec & $M_{200}$ & $r_{200}$ & SNR\\
& (deg) & (deg)& $({\rm h}^{-1} 10^{13} {\rm M}_{\odot})$ & $({\rm h}^{-1} {\rm Mpc})$ & ($r_{200}/\sigma_{r_{200}}$)\\
\hline
4-Cluster\\
A901a & 149.1099 & -9.9561 & $14.95^{+3.12}_{-4.32}$ & $1.11^{+0.07}_{-0.12}$ & 9.30 \\[5pt]
A901b & 148.9889 & -9.9841 &  $21.96^{+11.34}_{-10.15}$  & $1.26^{+0.19}_{-0.23}$ & 5.35 \\[5pt]
A902 & 149.1424 & -10.1666 & $2.78^{+1.45}_{-1.78}$ & $0.63^{+0.10}_{-0.18}$ & 3.47\\[5pt]
SW & 148.9101 & -10.1719 & $5.27^{+2.50}_{-2.52}$  & $0.78^{+0.11}_{-0.15}$ &  5.11\\[5pt]
\\
7-Cluster\\
A901a &  149.1099 &  -9.9561 &$12.60_{- 4.50}^{+ 3.37}$ & $1.05_{- 0.14}^{+ 0.09}$ & 7.29 \\[5pt]
A901$\alpha$ & 149.0943 &  -9.9208 & $1.90_{- 0.90}^{+ 0.95}$ & $0.56_{- 0.11}^{+ 0.08}$ & 5.19 \\[5pt]
A901b & 148.9889 & -9.9841 & $24.86_{- 10.48}^{+11.52}$  & $1.31_{- 0.22}^{+ 0.18}$ & 6.0 \\[5pt]
A902 & 149.1424 & -10.1666 &$ 2.96_{- 1.95}^{+ 1.43}$ &$0.65_{- 0.19}^{+ 0.09}$ & 3.33\\[5pt]
CB1 & 149.1650 & -10.1728 & $ 0.48_{- 0.43}^{+ 0.43}$ &$0.35_{- 0.18}^{+ 0.08}$ &  1.95\\[5pt]
SWa & 148.9240 & -10.1616 & $0.67_{- 0.61}^{+ 2.26}$ &$0.39_{- 0.22}^{+ 0.25}$ & 1.82\\[5pt]
SWb & 148.9070 & -10.1637 & $2.22_{- 1.0}^{+ 1.12}$ &$0.59_{- 0.10}^{+ 0.09}$ & 5.55\\[5pt]
\hline
\end{tabular}
\end{center}
\caption{Measurements of the virial radius and virial mass of the STAGES clusters, taken to be the mode and 1-$\sigma$ uncertainty on either side of the mode taken from the marginalised posterior distributions on $r_{200}$ for each cluster. RA and Dec label the centroid of each cluster considered, and are taken directly from H08. The ``4-cluster'' and ``7-cluster'' cases are chosen to mimic the analysis of H08, and to allow for easier comparison between masses derived using shear and magnification measurements, and centroid positions are taken from that analysis. For both the virial radius and virial mass, a flat prior on each is considered.}
\label{tab:NFW_res}
\end{table*}

\begin{figure}
\centering
\includegraphics[width = 0.48\textwidth]{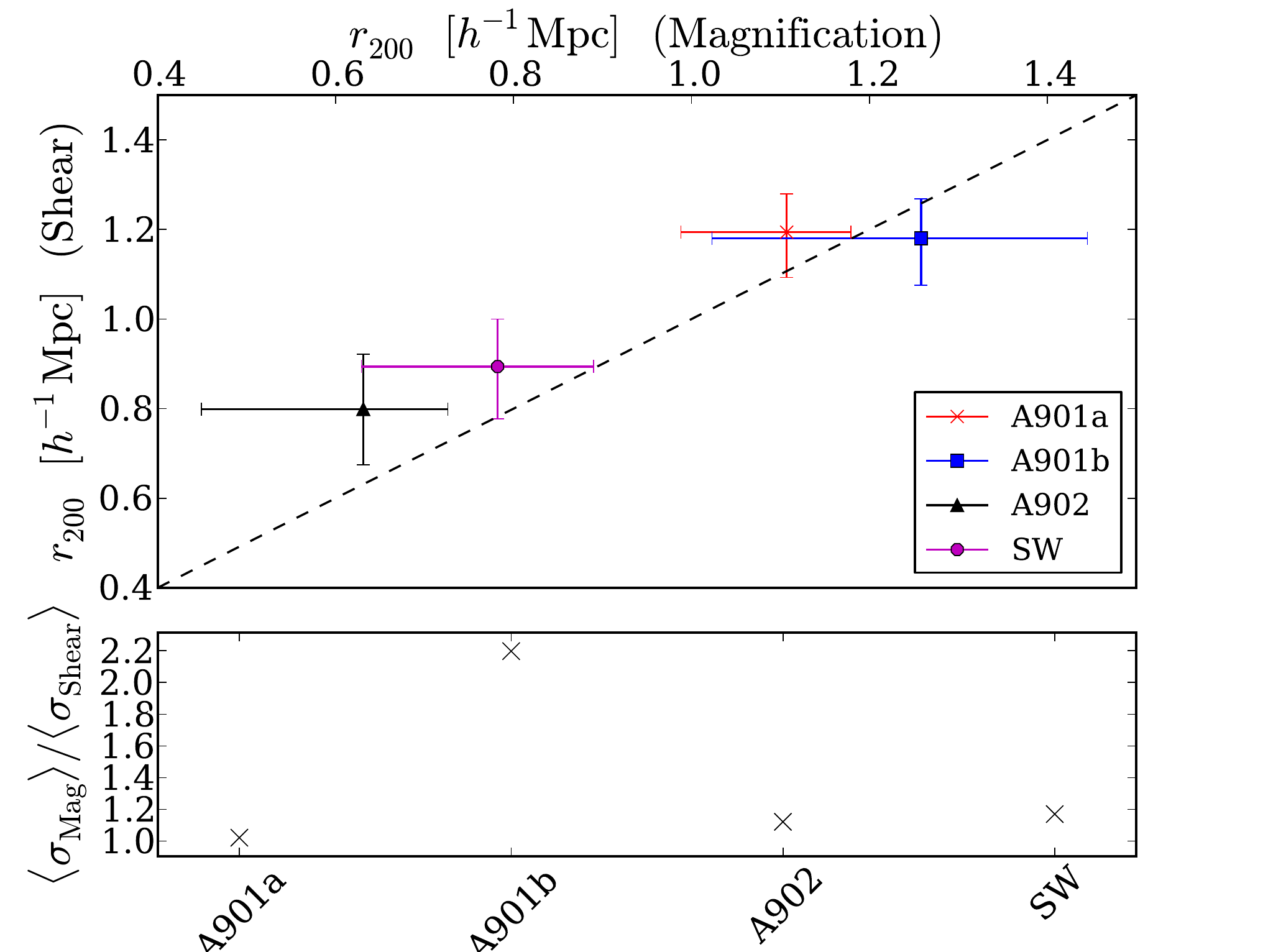}
\caption{Plot comparing the maximum-posterior estimates and uncertainties between the described size-magnitude magnification analysis and the shear analysis of H08, in terms of the recovered virial radius, for the one-halo case where the four main clusters (A901a, A901b, A902 and SW) are modelled on the field. {\it Top} shows the shear results on the ordinate axis, with the results of this investigation on the co-ordinate axis. The dashed diagonal line shows a one-to-one correspondence. {\it Bottom} shows the ratio of half the total 68\% confidence level for the magnification analysis to the shear result for each cluster.} \label{fig:Shear_Mag_Comp_4Cluster}
\end{figure}

Figure \ref{fig:Shear_Mag_Comp_4Cluster} shows the comparison between the maximum-likelihood estimate of the cluster virial radius from the shear analysis of H08 against the maximum-posterior results presented here. The top panel shows the recovered virial radius and $68\%$ confidence limit for each cluster, whilst the bottom panel shows the ratio of the total width of the 68\% confidence region in the magnification analysis to the shear analysis. We see good agreement between the magnification and shear results, however the magnification analysis typically produces lower virial radii than the shear analysis, particularly for A902 and the SW group, which represent the smallest modelled over-densities on the field in this case.

We see also that for A901a, A902 and SW the magnification estimate is produced with a comparable statistical uncertainty to the shear signal, with the magnification analysis producing estimates with a purely statistical uncertainty less than $20\%$ larger than the shear analysis in all three cases. This result is promising, particularly as one recalls due to limitations in the data we have applied core removal on all four clusters to reduce contamination by cluster members, thus reducing the source sample over that used for the shear analysis. Further, we do not have size measurements for the whole source sample. For A901b, the error on the magnification analysis is approximately twice as large as the equivalent-mass A901a, consistent with the reduction of the source sample around A901b through the application of a more strict faint magnitude cut (see section \ref{sec:Source_Selection}).

\begin{figure}
\centering
\includegraphics[width = 0.48\textwidth]{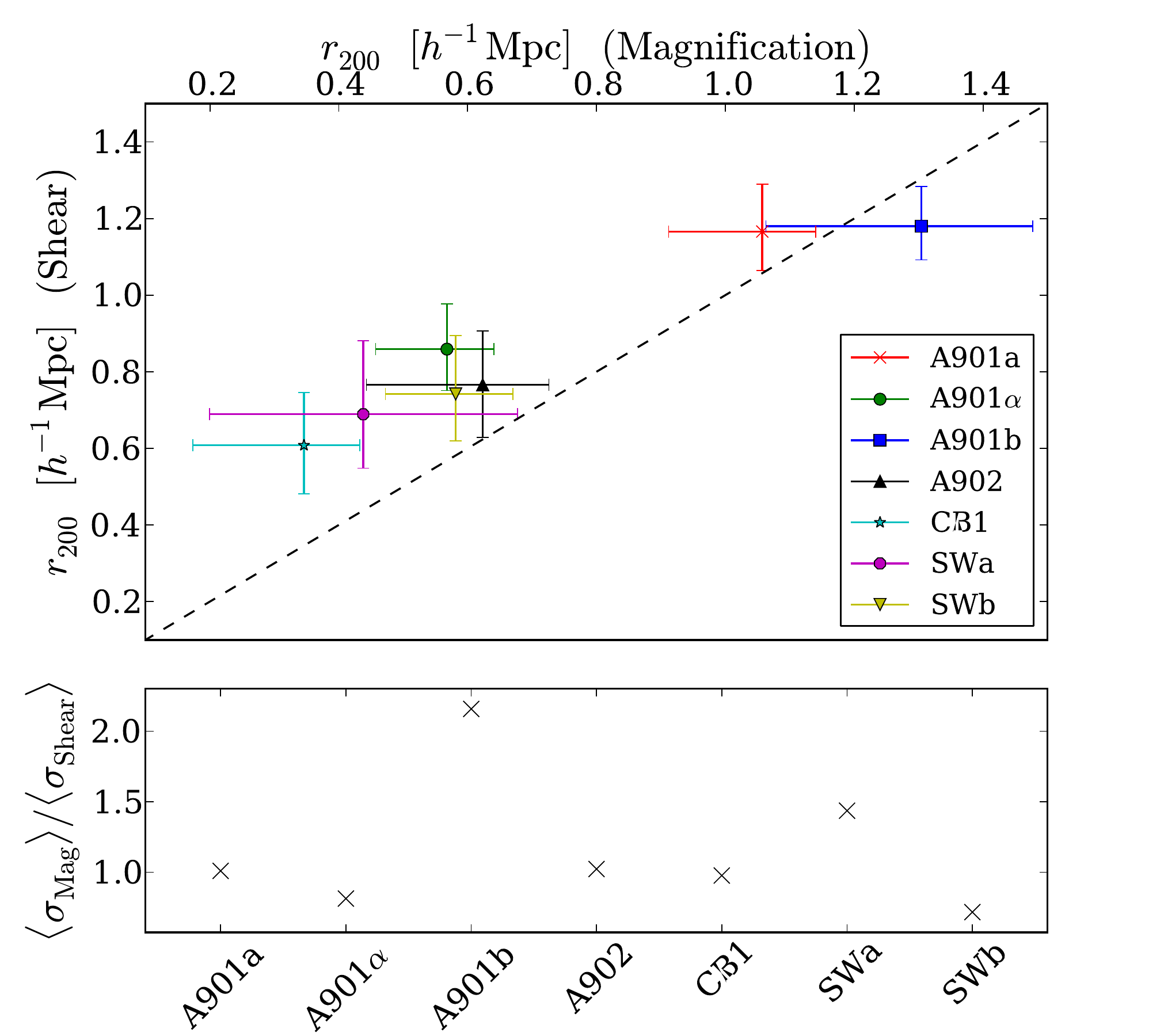}
\caption{As Figure \ref{fig:Shear_Mag_Comp_4Cluster}, but in the `7-cluster' case.} \label{fig:Shear_Mag_Comp_7Cluster}
\end{figure}

Figure \ref{fig:Shear_Mag_Comp_7Cluster} shows the same for the 7-cluster case. We see similar trends to the 4-cluster case, with the mode recovered virial radius from the magnification analysis typically lower than the shear results. For all clusters considered, the error on the virial radius estimate from the magnification is comparable to that of the shear, with the exception of A901b resulting from the stricter cuts used around this cluster.

\subsection{Comparison to other results}

In comparing the results presented here to the shear measurements, one must be aware of a few effects which can complicate such a comparison.
 Firstly, in Section \ref{sec:ApplicationToMocks} we have shown that the presence of measurement noise in the data may introduce a low bias which primarily affects the smallest structures, but is smaller than the expected error on recovered cluster parameters.
Secondly, In this analysis we have applied core subtraction on the source sample, which was not applied in the shear analysis. This can have multiple effects on the final result: as well as cluster member contamination such a subtraction should make the magnification analysis less susceptible to contamination of the signal by intrinsic size- and magnitude-density correlations by removing sources close to the cluster, however equivalent intrinsic ellipticity-density correlations may be present in the shear analysis; further, in both analyses it is assumed that the underlying dark matter mass profile is well described by an NFW profile with a fixed mass-concentration relation. If this is not true, then the subtraction of the core sample may introduce a discrepancy between the results through a model bias as the shear analysis is more sensitive to the core of the true lensing mass distribution than the core-subtracted magnification analysis.  
 Finally, in both cases the mass profile centre has been fixed to the values used in the shear analysis, which may introduce a centroid bias which will be more significant in the directional spin-2 shear analysis than the scalar magnification field \citep[][for a description on how mis-centering may affect each measurement]{2007arXiv0709.1159J}.

\cite{Ford:2014p2825} present measurements for a sample of 3D-Matched Filter clusters in the CFHTLenS survey using magnification bias (where the number density contrast of a distinct background source sample forms the estimator for the magnification field) and shear found that the magnification-derived cluster masses where systematically lower as a function of richness in comparison to shear mass measurement, similar to the trend we find here. In that analysis, the authors also find that the recovered mass of the magnification analysis is larger than the shear in the redshift range of the STAGES clusters, in contrast to the trend seen here, however it is noted that in that range their analysis may be affected by low-redshift contamination of the Lyman-break source sample. Such low redshift contamination may cause a positive bias due to the positive correlation between magnification due to an over-density and physical clustering. In the case presented here, the subtraction of sources around the BCG should limit the equivalent effect due to size- and magnitude-density correlations. The authors also discuss a range of other possible contaminants to the magnification signal which could cause the observed bias, including source obscuration, varying survey depth and seeing, galactic dust and stellar contamination. Source obscuration by cluster members \citep[see ][]{2015MNRAS.449.1259S} causes a reduction in the observed number density of distant sources, and is thus idiosyncratic to the study of magnification through source number density. Stellar contamination and noise in size and magnitude determination occurring from varying depth across the survey would require a correlation between these effects and cluster position, and is therefore not expected to be a source of bias in this analysis. Finally, systematic bias in size or magnitude measurements due to galactic dust are not expected to translate to bias in cluster parameters in this analysis due to the small are covered by the STAGES field, as systematic shifts across the field will affect both the source sample and the field sample in which the a-priori distributions are constructed. Thus whilst (with the exception of source obscuration) each effect has a counterpart in the type of analysis presented here, we do not expect that any will translate to significant bias in this analysis.

In \cite{Schmidt:2012p1106}, the authors presented a joint magnitude and size analysis on stacked groups in the COSMOS field, and found that the projected surface mass density from the magnification analysis was consistent with a shear analysis within the uncertainties, but with a signal--to--noise approximately $40\%$ of the shear value. In this analysis, we also find that the magnification analysis returns the cluster mass to a typically lower signal--to--noise than the shear, however the reduction here less severe (with magnification signal--to--noise ranging from $\sim 53\%$ of the shear equivalent for A901b to $80\%$ for A901a), and primarily driven by a lower recovered cluster mass in the magnification measurement, rather than driven by statistical uncertainties except in the case of A901b where particularly conservative cuts are enforced. We note that in \cite{Schmidt:2012p1106}, the authors used quadrupole measures to determine the size of their source galaxies. In Appendix \ref{sec:STAGES_RRG_Measurement}, we find that the use of such a measure is complicated by the application of a weight function which introduces systematic bias in the size measure as a function of PSF and source ellipticity is un-corrected for. Further, we find that quadrupole moment based measures of size are unable to distinguish between large, low surface-brightness sources and small, large surface-brightness sources except for high signal--to--noise images. As a result, we conclude that the use of such a measure is likely to introduce inaccuracies in the recovered size measure for the smallest of faintest sources. Whilst the application of a small source cut may limit the effect of the PSF on the recovered size, the dependence on ellipticity and intrinsic surface brightness will remain, which are likely to introduce noise to the source sample of galaxies. In such a case, the majority of the information may still be provided by the magnitude estimation, thus limiting the impact of the inaccurate size measure in the form of a bias, but with an increasing statistical error. 
Finally, we note that the application of \cite{Schmidt:2012p1106} assumes a multivariate Gaussian in log-size and magnitude for the underlying size-magnitude distribution. Whilst we find the log-size is approximately Gaussian in this application (see Figure \ref{fig:SizeMag_Distribution_Master}), this is not the case for the magnitude distribution. Enforcing such an assumption is likely to introduce a bias in the recovered cluster mass, however the level of such a bias is non-trivial to quantify. In the method motivated here, such an assumption can be avoided with an appropriate model for the a prior intrinsic size-magnitude distribution, or if this distribution is measured directly from the data, as in this application.

In \cite{Alsing:2014p2846} the authors consider the application of a similar Bayesian size-magnification inference on CFHTLenS data, and find that the convergence field can be recovered with uncertainty $\sigma_\kappa \sim 0.8$, compared to $\sigma_e \sim 0.4$ for the ellipticity distribution. The authors then investigate the ability of a size-magnitude analysis to provide forecast constraints on cosmological parameters through the use of convergence power spectra with shot noise contribution determined by this value. They find that magnification alone is less powerful than shear, but that the addition of magnification to a shear analysis can provide valuable additional information, particularly in the presence of shear systematics which must be taken into account with flexible models whose model parameters are then marginalised over. Whilst the shear and convergence share the same second-order statistics, they probe the mass distribution in subtly different ways: the shear is sensitive to the differential mass profile, whilst the magnification is a direct probe of the local mass distribution. As such, it is not obvious that the reduced constraining power of the magnification analysis in cosmological situations mirrors exactly its ability at direct mass estimates. In \cite{Rozo:2010p1496} the authors forecast the ability of an ellipticity, size or number density analysis to probe the mass-concentration plane and find that size measurements produced tighter constraints on mass than shear alone. That analysis does not take into account the differing number density between the size and shear sample, as is the case here, nor a joint size-magnitude analysis, but the seeming equivalence of the shear and size signals agrees with the trend we see here, and this application is supportive of those results.

\section{Conclusions}\label{Sec:Conclusions}

In this paper, we have demonstrated the use of a joint size and flux/magnitude magnification analysis as a probe of the dark matter profile of a single lens. To do so, we have used a Bayesian formalism which allows one to produce a posterior probability distribution on lensing mass distribution model parameters for each individual source cluster, and which can be combined to give a joint distribution using the full source sample. To do so, one must have a priori knowledge of the intrinsic size and magnitude distributions of the source sample. Whilst this is not directly possible, we argue that one can acquire this information directly from the data by considering a source sample across the whole field, provided the average magnification is unity across the field. The method allows for the natural inclusion of a redshift distribution for sources whose redshift is not known, as well as a natural method of accounting for cluster members in the source sample, and intrinsic size-magnitude correlations, as well as measurement uncertainty in the source size and magnitude.

By applying the method to mock catalogues, we showed that the method can give unbiased mass estimates across a range of masses provided that the size and magnitudes of the source sample are well-measured. We argue by comparison with the measured shear values on the STAGES field that the size-magnitude analysis could provide competitive constraints on the cluster mass, in the idealised case where sizes are known for the full sample, measurements are exact and no additional source cuts must be used, however we note that in the application to the data we must account for the fact that these simplifications no longer hold.

We find that the method is robust to a variety of possible systematics, but note that noise bias resulting from uncertainty in the measurement of source size and magnitude may produce a significant low bias in the recovered lens mass if not accounted for. Whilst the inclusion of a method to account for this uncertainty is straightforward theoretically, it is restricted by computational limitations in the current analysis. 

We applied the method to the STAGES data, and produce posterior distributions on the cluster virial radius for the four main structures on the STAGES field. We find that the magnification analysis provided a detection of A901a, A901b, A902, and the SW group to a signal--to--noise of $9.3, 5.4, 3.5$ and $5.1$ when reported in terms of the virial radius. This compares well with the shear signal--to--noise for the same clusters, with the magnification analysis giving a single--to--noise ranging from 64\% (for A901b) to 80\% (for A901a) of the shear result, and largely driven by the lower recovered virial radius values. When the SW group is split into two over-densities, and additional over-density around A901a (named A901$\alpha$) and the background cluster of CB1 is modelled, as motivated in H08, we find that A901a, A901$\alpha$, A901b, A902, and SWb are detected to a signal--to--noise ratio of $7.3, 5.1, 5.4, 3.5, 5.3$ respectively, whilst CB1 and SWa have a maximum-posterior which is non-zero to a $2$ and $1.9\sigma$. In this case, the signal--to--noise ratio of the magnification analysis ranges from 45\% of the shear result for CB1, to 110\% for SWb, with A901a, A901$\alpha$, and A902 giving 77, 73 and 63\% respectively. We find that the statistical uncertainty on cluster mass for considered clusters is comparable between the shear and magnification analyses, with the exception of A901b where a strict faint magnitude cut must be applied to ensure the accuracy of the measurement, and that the reduction in signal--to--noise in the magnification analysis is driven instead by a low recovered virial radius for the majority of the clusters. Accounting for the fact that a core subtraction was necessitated for the magnification analysis to limit contamination by cluster members and the effect of intrinsic size- and magnitude-density correlations, thereby significantly reducing the size of the source sample, we conclude that the magnification analysis provides a competitive way to constrain lens mass profiles, with the caveat that the lower recovered values compared to shear must be better understood in the future.

As we move to larger and more expensive surveys, with progressively more stringent science requirements, it will become increasingly important to use the full range of information available to us to produce scientific results. For lensing surveys where shear analysis is already de rigueur, this can be easily achieved using magnification as a probe, where the size, magnitude or number density measurements required for a magnification analysis are already produced as an off-shoot of the main science drivers. With the burgeoning list of investigations which show that there is vital information in the magnification signal in a cosmological context \citep{Duncan:2014p2569,Alsing:2014p2846,Eifler:2013p2722,Gaztanaga:2012p1194,Eriksen:2015p2849} and in lens reconstruction \citep{Rozo:2010p1496,Schmidt:2012p1106,Ford:2014p2751,Ford:2014p2825,Bauer:2011p2066,Umetsu:2014p2726,Hildebrandt:2011p2755}, it is more clear than ever that time spent developing the means to use this information, through producing accurate size and magnitude measurements or modelling systematics, will be well spent.

\section*{Acknowledgments}

CAJD would like to thank Graham Smith, Boris H{\"a}u{\ss}ler, Ami Choi and Justin Alsing for useful discussions, and the anonymous referee for their helpful comments on this paper. CH and CAJD acknowledge support from the European Research Council under grant number 240185. BJ is supported by an STFC Ernest Rutherford Fellowship, grant reference ST/J004421/1. This paper uses data from the HST-STAGES program number GO-10395.  We thank the COMBO-17 and STAGES team (PI Meghan Gray) for making this data set public.

\bibliographystyle{mn2e}
\setlength{\bibhang}{2.0em}
\setlength\labelwidth{0.0em}
\bibliography{Papers.bib}

\appendix

\section{Application to quadrupole sizes}\label{sec:STAGES_RRG_Measurement}

In the main body of this text, we use the GALFIT size measures from the publicly available STAGES source catalogue. In this appendix, we investigate the use of an alternative quadrupole moment-based estimator, as used in \cite{Schmidt:2012p1106}, in order  to measure the galaxy size for every source with a quadrupole-based measurement of shear. If one could measure galaxy size for each source with a quadrupole shear estimate in the H08 catalogue, one would increase the size of the source size sample and therefore minimise statistical uncertainty in the recovered cluster profile parameters. As part of this analysis, we investigated the use of such a measure as part of the analysis, and found complications in its use. In this section, we present an investigation into the use of quadrupole size measures, applying the PSF correction of \cite{Rhodes:2000p2068} (hereafter RRG) to multiple runs of the GREAT 10 image simulation suites, for a range of input Sersic scale radii and signal--to--noise ratio. The PSF is modelled as an isotropic Moffat profile, with width $\sigma_{\rm PSF} = 3.3$pix, corresponding to the isotropic width of the PSF measured on the STAGES field through the measurement of stellar images. Galaxy images are taken to be randomly orientated, with an ellipticity sampled from the ellipticity distribution of \cite{Miller:2013p2259}. 

Using quadrupole moments, galaxy size is determined through combinations of the quadrupole moment, defined as the integral of the weighted surface brightness profile of the image
\be\label{eqn:QuadMoment_SBWeight}
J_{ij} = \frac{\int {\rm d}^2\theta \;\theta_i \theta_j W[I(\bm{\theta})]I(\bm{\theta})}{\int {\rm d}^2\theta \;W[I(\theta)]I(\theta)}
\ee
where $W[I(\theta)]$ is a window function, normalised to unity over all space whose inclusion ensures convergence of the integral over noisy images, or where galaxies are not isolated on the image, and for convenience of notation, we have defined the origin of the angle $\theta$ from the centroid of the image. Following RRG, the window function is chosen to be a Gaussian, whose width is set by the measured Source Extractor \citep{Bertin:1996p2806}  flux radius of the source. Source size can then be defined as 
\bea
S_1 &=& \det(J)^{\frac{1}{4}} = (J_{11}J_{22} - J_{12}^2)^{\frac{1}{4}} \label{eqn:quadrupole_SizeMeasurement_detJ}\\
S_2 &=& (J_{11}+J_{22})^{\frac{1}{2}}, \label{eqn:quadrupole_SizeMeasurement_TrJ}
\eea
so both definitions have the units of {\it length}. Under the action of a foreground lens, it can be shown that each size measure is transformed as
\bea
&S_1& = \left[\frac{J^s_{11}J^s_{22} - (J_{12}^s)^2}{\left[(1-\kappa)^2-|\gamma|^2\right]^2}\right]^\frac{1}{4} = \mu^{\frac{1}{2}}S_1^s, \\
&S_2& = [(J^s_{11}+J^s_{22})[1+2\kappa] + 2(J^s_{11}-J^s_{22})\gamma_1 + 4J^s_{12}\gamma_2]^{\frac{1}{2}}. \;\;\;
\eea
Thus, the transformation of $S_1$ is exact for a noiseless image, whereas $S_2$ only transforms according to the standard lensing equation (\ref{eqn:Lensing_Relations__Size}) when the weak lensing limit is enforced. As a result, the size measure of $S_2$ can be expected to be biased for those sources chosen near the centre of the cluster, where the weak lensing limit is least applicable. Whilst $S_1$ transforms exactly, the measurement of size using this definition is noisier due to the non-linear combination of quadrupole moments, complicating the following application of calibration on galaxy size using image simulations. Consequently, for the remainder of this text, we will use the size measure given as $S_2$, will frequently be referred to using the label `{\rm Tr}(J)'. 

The use of the RRG correction to the measured quadrupole moments on the field image allows for the determination of a source size which has been corrected for the effects of the PSF and optical distortion of the telescope, however the method provides no means for the correction of the image due to the use of a weight function. The application of such a weight function down-weights the noisy surface brightness profile towards the wings, and as such the measured size using such a quadrupole moment is dependent on the choice of the weight function width when carrying out the measurement. As typical applications of such measures in source ellipticity determination take the weight function width to be an initial guess of the source size (such as Source Extractor flux radius), the quadrupole determined size is dependent on the accuracy of the initial guess, and in this case the ability of SExtractor itself to accurately measure source size: as such, even though the RRG method provide a means of correcting the moments for the measured PSF, the moments themselves may still be affected by the PSF through the use of the uncorrected flux radius to set the weight width, particularly for the intrinsically smallest bodies. In addition there will be biases in SExtractor-derived sizes due to the source ellipticity and low signal--to--noise ratio. The calibration of measured sizes must therefore initially correct for the use of the weight function. 

In the absence of a mathematically motivated correction for the weight function, we used an initial empirical calibration, measured from high signal--to--noise simulated images. In this application, by sampling the absolute ellipticity from the distribution of \cite{Miller:2013p2259}, we implicitly marginalise over an intrinsic ellipticity distribution and thus the results include the effect of implicit bias in measured size due to the use of biased SExtractor flux-radius initial guess. As such, these results will hold where the underlying ellipticity distribution is given by that of \cite{Miller:2013p2259}, however the calibration may not be exact where the underlying distribution is different. For the conclusions presented here, this assumption is enough to determine trends, but care would be needed in the application to data. Further, it is also clear that this effect is likely to induce a size-ellipticity correlation which must be taken into account where the full size-magnitude-ellipticity analysis detailed in the main text is used, and a such one can already surmise that the quadrupole-based measures complicate such an application.

\begin{figure}
\centering
\includegraphics[width = 0.5\textwidth]{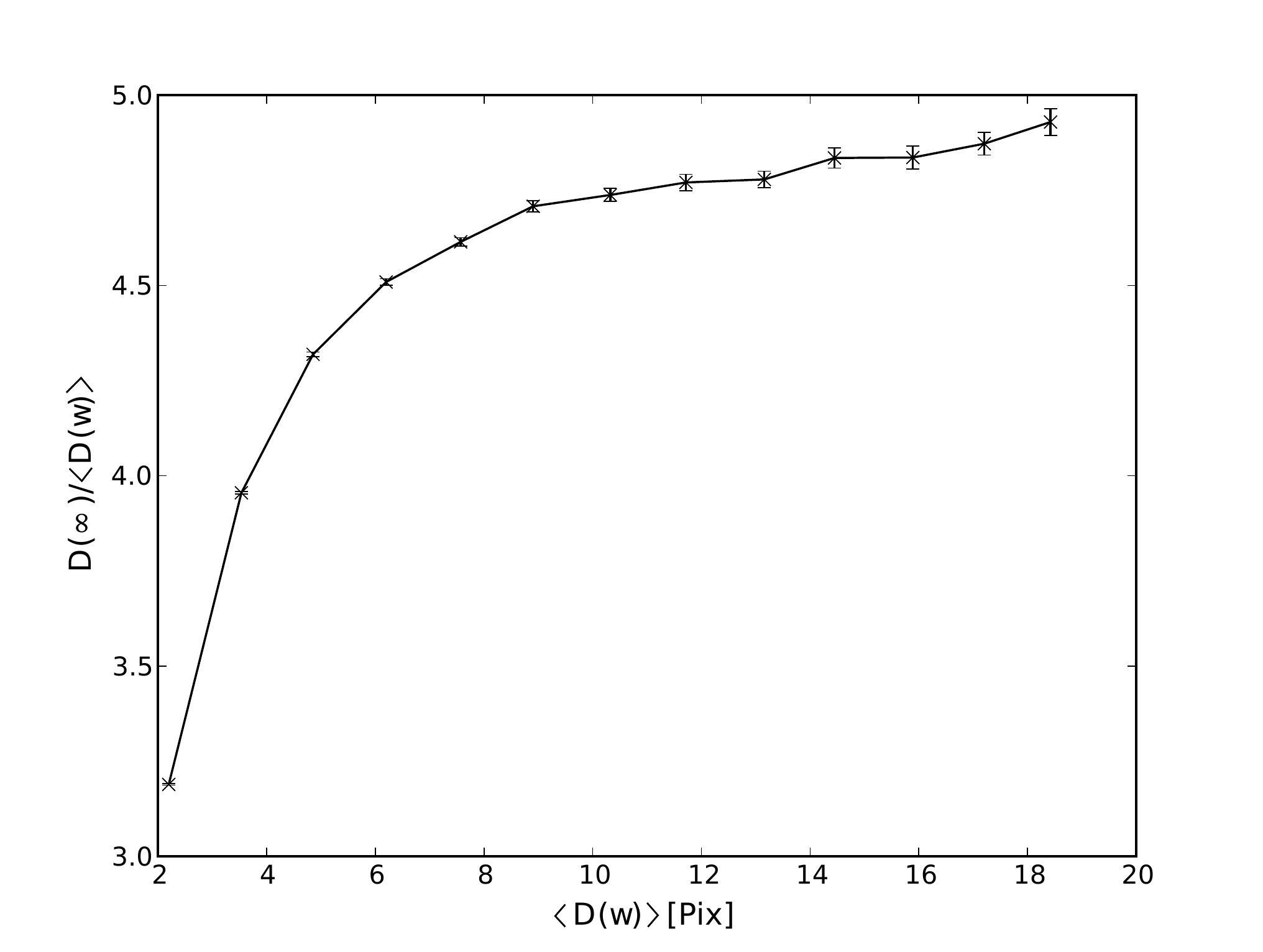}
\caption{Calibration of measured source sizes, measured from high signal--to--noise ratio image simulations, to account for the sensitivity of RRG-derived sizes to the width of the Gaussian weight function used, in this case taken to be the flux radius, as measured by SExtractor}\label{Fig:WCalib}
\end{figure}

A high signal--to--noise image can be calibrated to an equivalent `unweighted' size measurement as $D \to GD(w)$, where G = $D(\infty)/\langle D(w) \rangle$, and $D(w)$ here is used to label quadrupole measured size using a gaussian weight function with width $w$. $\langle D(w) \rangle$ is measured from image simulations at high flux signal--to--noise (SNR $= 200$), whilst $D(\infty)$ is calculated analytically in the noise-free case of a circularly symmetric Sersic profile. Figure \ref{Fig:WCalib} shows the ratio of unweighted quadrupole size to the average measured size for a set of simulated galaxy images as a function of measured source size. One can see that the ratio of unweighted size to measured size decreases quickly as $D(w) \to 0$, whilst the ratio becomes linear for larger sizes. The decrease at small sizes results from the effect of the PSF on the measured SExtractor flux radius: since the flux radius is not corrected for the PSF, the PSF causes the measured flux radius for the smallest sources to be biased high. Consequently, the weight function applied to these galaxies has a respectively larger width for these sources than for those much larger than the PSF, resulting in a measured quadrupole size which is systematically larger due solely to the choice of setting the weight function width to be the flux radius. 

It is important to note that it is the change in the correction with measured size which is important in the application of this correction: if a flat relation was observed for all weighted sizes, the resulting a priori intrinsic size distribution and source sample would have their measured size shifted by the same amount, causing no qualitative change in the measured size distribution nor the measured magnification factor. Is is worth noting, however, that even at larger weighted sizes, the corrective factor is not flat: even without the effect of the PSF on the weight function width, larger galaxies would require a relatively larger correction to their size than smaller galaxies which is likely to change the properties and statistics between the corrected and uncorrected size distributions. For sources whose measured weighted size is larger than the range considered here, the correction factor is taken from linear extrapolation.

\begin{figure*}
\centering
\includegraphics[width = \textwidth]{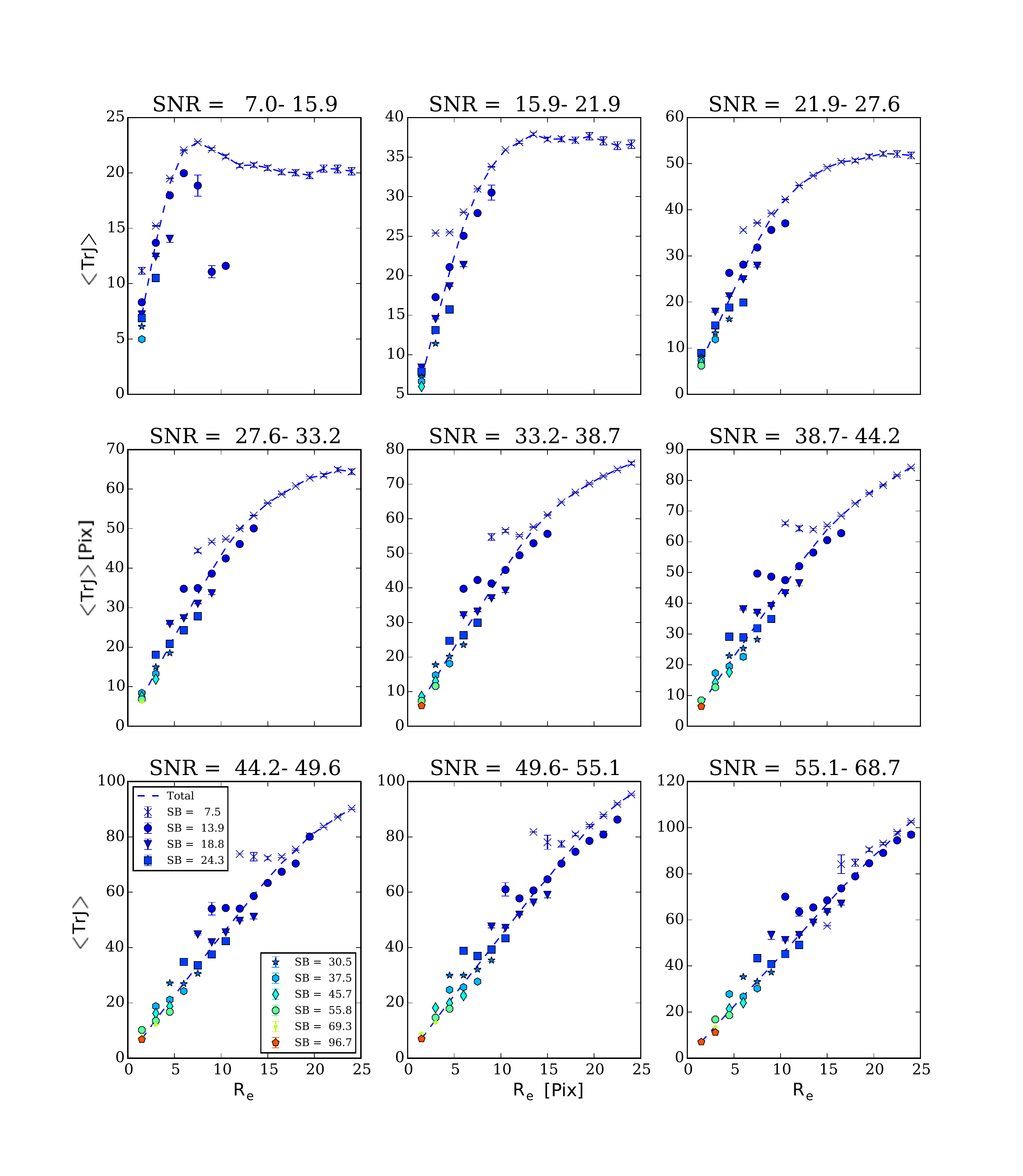}
\caption{Plot of quadrupole measured size against input size, binned by measured signal--to--noise--ratio (SNR). Each panel corresponds to a given signal--to--noise bin, the ordinate axis gives the mean measured size using quadrupole moments for simulated Sersic galaxies with scale radius given by the co-ordinate axis. Coloured points correspond to the measurement in bins of intrinsic surface brightness, whilst the blue dashed line gives the combination of all surface brightness bins. All sizes are given units of pixel size, and surface brightness (SB) is given in units of counts per pixel squared.}\label{Fig:SSNRCalib}
\end{figure*}

Figure \ref{Fig:SSNRCalib} shows the measured quadrupole size for a series of input Sersic scale radii and measured signal--to--noise bins, where the measured signal--to--noise is taken as the ratio of SExtractor FLUX\_BEST to its equivalent error, colour coded by intrinsic surface brightness. The measured size has been corrected using the above window function calibration. One can see that for the range of scale radii considered here, the relation between the measured size and input size is linear where the signal--to--noise is large, suggesting that for high signal--to--noise sources the quadrupole size is unbiased with respect to the intrinsic size of the source. It is worth noting that even at high signal--to--noise, the respective increase in measured size due to effect of the PSF on weight function width is not evident in these plots, suggesting that the application of the first level of calibration has successfully accounted for this trend. For low signal--to--noise sources, one can see that the quadrupole measured size does not follow the input size linearly, with a turnover seen for the largest simulated galaxies in that bin. As the measured signal--to--noise increases, this turnover is pushed to larger values of input size and recovered size, to the point where it is no longer observable on the input scales seen here. Conversely, one can see that the point of turnover trends to smaller sizes with decreasing signal--to--noise ratio. 

This turnover results from the fact that one observes only the tip of the surface brightness profile for the largest sources above the noise: these sources are intrinsically large and are observed as faint, with a low surface brightness. The wings of the surface brightness profile for an intrinsically large galaxy fall below the noise level of the image, and the observed boundary containing a given fraction of the total flux of the noisy image is smaller than in the noise-free case, causing a systematic underestimation of the source's size. As a result, the relation between measured size and input size is non-monotonic, and it becomes impossible to distinguish between an intrinsically small galaxy, and an intrinsically larger body for which only the central section of the profile is observed above poisson noise. This affects the RRG measured size in two ways: first, the SExtractor flux radius is underestimated causing a respective decrease in the width of the weight function used; and secondly, beyond the point where the galaxy surface brightness profile is sub-dominant to the background, any addition to the quadrupole moment is noise dominated. Since the quadrupole moment integrates beyond the measured flux radius, the down-turn is less pronounced than observed in the flux radius measurement itself (not shown here).

Figure \ref{Fig:SSNRCalib} therefore suggests that size measurements using quadrupole moments are not reliable at low signal--to-noise, where the method cannot distinguish between faint, large sources and bright, small sources. At larger signal--to--noise, where the tunrover is pushed to larger instrinsic sizes, the distinction is reinforced by the reduced probability of observing a source with such a large intrinsic size. It is worth noting that Figure \ref{Fig:SSNRCalib} suggests that in part this degeneracy can be alleviated by implementing a cut on surface brightness, thereby removing the low surface brightness sources at low signal--to--noise, however the measured surface brightness is itself affected by the same effect, complicating the effective removal of these sources from the sample.

This investigation suggests that the use of quadrupole moments is inappropriate for accurate size measures without complicated calibration on realistic image simulations. In contrast, the use of model-fitting methods like GALFIT will avoid most of these problems, as the model may still be fitted to the peak of the surface brightness profile over the background, and thereby giving information of the profile out to the wings. Further, such a method will not require the use of a complicated weighting function which is itself a function of the measured size, and will therefore not need calibration to remove the window function (and secondary effects such as PSF and ellipticity bias), as done here. In \cite{2007ApJS..172..615H}, the authors show through the application to image simulations that GALFIT size measures using the sky estimates from the GALAPAGOS routine recover the galaxy size well over a wide range of magnitudes, with only a small deviation in the magnitude ranges considered in this analysis.

\section{Alternatives to source selection criteria}\label{Sec:Alternative-Source-Selection}

In section \ref{sec:Source_Selection}, we detailed the use of magnitude, redshift and core cuts as a means of limiting the impact of cluster member contamination of the source sample and ensuring the accuracy of the individual source measures. This section couched the discussion on the accuracy of the results in a frequentist way, using discussion of possible bias in the recovered cluster profile parameters. In that sense, the bias instead can be interpreted as an acknowledgement of the limitations of the forward modelling process used. In this application, we try to clean the data to fit the model used in the analysis (that is, one that does not account for the presence of such contaminants), however an alternative approach is to use a more realistic model which attempts to account for these systematics, simplifying the interpretation of results in a physical way and maximising the available source sample.

An extension to the applied method has already been discussed in the preceding sections to account for uncertainty in individual source size and magnitude measurements, by integrating over a latent variable which describes noisy estimators of the intrinsic values of  these quantities. The application of such a method is complicated by the need to know  the relation between the estimate and the intrinsic quantity, which must account not only for pure statistical errors on the measurement, but also systematic uncertainty due to limitations of the data or measurement itself, for example through subject blending, foreground masking, or ability to only measure the peak of the surface brightness profile in faint sources (see Appendix \ref{sec:STAGES_RRG_Measurement}).

In \cite{Velander:2010p636} the effect of cluster contamination is mollified by weighting close lens-source pairs according to their assigned lensing efficiency, taken as the mean for that source sampled from a expected redshift distribution. Such a weighting would reduce the contribution from sources expected to be radially close to the lens. The implementation of such a weighting is non-trivial in the analysis we have presented here, without reducing the measurement from individual source posteriors to statistics.

A natural method to include the presence of un-lensed cluster members is the sample would be to edit the a priori redshift distribution to include the presence of a fraction of cluster contaminants. The redshift distribution chosen in this application is expected to accurately represent the distribution of field galaxies, but does not account for the local over-density of galaxies at a certain angular position and redshift due the presence of a cluster at that position. The distribution could be made to better represent the source sample including cluster members by the addition of a spike in the redshift distribution, centred on the mean redshift of the cluster members and with width representative of the uncertainty in the photometric redshifts at that redshift. Such a modification would only account for cluster contamination of this type provided the size-magnitude distribution for the cluster members is accurately described by that measured for the field sample, and provided any size- or magnitude-density correlations were small. If the former assumption does not hold, the a priori size-magnitude distribution would need to account for whether the source is a field galaxy or a cluster member: since such a distinction is not possible in this case, such a situation could not be easily rectified. If the latter assumption did not hold, the method could be generalised to include a magnitude- or size-density correlation through a modification to the relationship between the observed and intrinsic sizes and magnitudes, given in equation \ref{eqn:Observed_Intrinsic_SizeMagnitude_Relation} where the relation presented is assumed to be due to the magnification effect only. Similarly, the modification to the redshift distribution would need to be spatially varying if the sources are selected in a region on which the total projected number density of cluster members varies on a length scale which is shorter than the source selection region.

The use of such techniques may allow for the use of less stringent core cuts, maximising the sample size and consequently minimising the statistical noise of the analysis. The investigation of this is left to future work.

\bsp

\label{lastpage}

\end{document}